\documentclass[12pt]{article}
\usepackage{amssymb,amsmath,epsfig}
\usepackage{cite}
 \allowdisplaybreaks

\begin{document}
\title{\bf The Measure of Complexity in Charged Celestial Bodies in $f(R,T,R_{\mu\nu}T^{\mu\nu})$ Gravity}
\author{Z. Yousaf$^1$ \thanks{zeeshan.math@pu.edu.pk}, M. Z. Bhatti$^1$ \thanks{mzaeem.math@pu.edu.pk},
T. Naseer$^1$ \thanks{tayyabnaseer48@yahoo.com} and I. Ahmad$^2$ \thanks{irfan.pu6@gmail.com}\\
$^1$Department of Mathematics, University of the Punjab,\\
Quaid-i-Azam Campus, Lahore-54590, Pakistan\\
$^2$ Department of Mathematics and Statistics,\\
The University of Lahore, Lahore, Pakistan.}

\date{}

\maketitle
\begin{abstract}
In this paper, we investigate irregularities in a cylindrical
self-gravitating system which contains the properties of an
imperfect matter and electromagnetic field. For $f(R,T,Q)$ theory,
in which $R$ represents the Ricci scalar and $T$ shows the trace of
matter stress-energy tensor while $Q\equiv
R_{\gamma\delta}T^{\gamma\delta}$, the field equations containing
electric charge, mass functions and Darmois junction conditions at
the hypersurface are examined. We have adopted new definition of
complexity introduced by Herrera \cite{herrera2018new}, generalized
it for the static charged cylindrically symmetric case in $f(R,T,Q)$
theory by performing a detailed analysis on the orthogonal splitting
of the Riemann curvature tensor. One of the effective scalars,
$Y_{TF}$, has been recognized as a complexity factor. This factor is
comprised of certain physical components of the fluid such as
irregularity in energy density, locally pressure anisotropy and
electric charge (arranged in a specific way). In addition, the
effects of extra curvature terms of modified gravity are examined by
making the relations among the complexity factor, Weyl scalar and
Tolman mass.
\end{abstract}

\section{Introduction}

General Relativity (GR) is an extensively accepted theory that was
proposed by Albert Einstein in 1915 in which he related matter and
spacetime with the help of Einstein field equations. In accordance
with his beliefs, there does not occur any contraction or expansion
in our universe since from the big bang. After a short time, the
red-shifted light has been received from some galaxies when Edwin
Hubble was trying to measure the velocities of those galaxies in
1929, and thus he concluded that our universe is in state of
expansion. Some fascinating discoveries have recently shown that our
cosmic dynamics needs a deeper understanding due to very high amount
(almost 95 percent) of dark matter and dark energy in our cosmos.
Many astrophysicists have expressed their interest in exploring
various ways of explaining the influence of our cosmic dark source
components. A number of modified gravitational theories have been
suggested recently which are significant for the study of dynamical
features of cosmic expansion. By substituting the generic function
of the Ricci scalar $R$ in place of $R$ in an action, the
straightforward generalization of GR has been obtained which is
known as $f(R)$ theory (for details on modified gravity please see
\cite{copeland2006dynamics,
nojiri2011unified,nojiri2017modified,capozziello2010beyond,capozziello2011extended,
de2010f,joyce2015beyond,cai2016f,bamba2015inflationary,yousaf2016influence,yousaf2016causes}).
To investigate the mysterious aspects of our cosmos, Nojiri and
Odintsov \cite{nojiri2007introduction} affirmed the consistency of
$f(R)$ theory by designing different models in this theory that are
compatible with certain solar system tests. In $f(R)$ theory, Bamba
\emph{et al.} \cite{bamba2012dark} discussed the features of several
cosmologies containing dark energy and $\Lambda$CDM-like universe by
considering different models. Different factors that could produce
irregularities in any self-gravitating object have been examined by
various researchers
\cite{konoplich1999formation,yousaf2019non,yousaf2019tilted,bhatti2017gravitational,sharif2019complexity,doi:10.1142/S0217732319503334,yadav2020existence},
and by taking some applications into consideration, they analyzed
rigorous solutions to the field equations in $f(R)$ theory. Yousaf
and his collaborators
\cite{yousaf2016electromagnetic,bhatti2017stabilitya,yousaf2020construction,bhatti2020stability,yousaf2020gravastars}
concluded that for some particular modified theory, the
gravitational models are quite appropriate for hosting compact stars
with smaller radii but having larger masses.

Harko \emph{et al.} \cite{harko2011f} first studied the effects of arbitrary coupling between matter and curvature by introducing $f(R,T)$ theory, where $T$ is the trace of stress-energy tensor. For this theory, they investigated the impact of additional force term on moving particles, and field equations are also evaluated with the help of variational principle. By considering some models in $f(R,T)$ theory, Baffou \emph{et al.} \cite{baffou2015cosmological} found their stability conditions, and established several cosmological solutions that suit observations. The more generalized form of $f(R,T)$ theory is known as $f(R,T,Q)$ theory, which was suggested by Haghani \emph{et al.} \cite{haghani2013further} and it incorporates the matter lagrangian consisting of the strong interaction between curvature and fluid. The equations of motion were also calculated by assuming Lagrange multiplier method in this theory. Odintsov and S\'{a}ez-G\'{o}mez \cite{odintsov2013f} noticed that the spacetime is strongly non-minimally coupled with geometry and found that the cosmic expansion may be the result of extra curvature terms of $f(R,T,Q)$ gravity. Ayuso \emph{et al.} \cite{ayuso2015consistency} inspected the stability conditions in $f(R,T,Q)$ gravity for certain specific scalar and vector fields, and also calculated higher order field equations. Baffou \emph{et al.} \cite{baffou2016exploring} analyzed the consistency criteria for $f(R,T,Q)$ theory via some particular solutions and claimed that one can better understand the early stages of our cosmic expansion through modified correction terms. In addition, by using computational techniques, they explored the stability of certain models in this theory. The gravitational collapse of a self-gravitating object was studied by Bhatti and his colleagues \cite{yousaf2016stability,Yousaf2017,yousaf2017stability,bhatti2019dissipative}, and some relations between physical variables containing in field equations and the Conformal tensor are also found.

A complexity can be defined as a combination of several components which can lead a stable self-gravitating system to be complicated. Here, the existing results in GR for zero complexity must be explained under the influence of extra curvature terms of $f(R,T,Q)$ theory. In various fields of science, different approaches to define a complexity can be observed. L\'{o}pez-Ruiz \emph{et al.} \cite{lopez1995statistical,calbet2001tendency,catalan2002features} defined this term by entropy and information amongst all those factors. Entropy of any system can be determined by measuring its disorderness, whereas the knowledge that one can gain by studying a system could be called as information. L\'{o}pez-Ruiz \emph{et al.} \cite{lopez1995statistical} then developed this definition by means of a term, i.e., disequilibrium.

The term complexity can be demonstrated in physics by assuming simplest structures, which by definition do not have any complications, such as isolated ideal gas and perfect crystal. By theirs very definition, such both structures are extreme in all aspects. The first system is entirely unstable in nature as it consists of randomly moving particles, and therefore gives maximum possible information because all particles have equal involvement in this system. In second system, the elements that construct this structure are organized in an orderly manner and therefore it provides minimum data set because to known its nature, it is sufficient to analyze its small portion. It can thus be seen in the former case that there is maximum disequilibrium, while in the later case it is found to be zero. The notion of complexity factor can also be used in astrophysics to examine the structural features of self-gravitating (cylindrical or spherical) systems. Usually, there can be various factors in any system such as pressure, heat dissipation and energy density which contribute in making the system more complex. The complexity of any system cannot be measured via its energy density only, unless the pressure component in the stress-energy tensor is taken into account.

Usually for cylindrical objects, the nature of several physical characteristics was analyzed at different scales. In particular, the phenomenon which encourage to consider the cylindrical symmetry are gravitational collapse and its radiations, rotating fluids and spinning massive structures. By definition of Birkhoff's theorem, outside a spherically self-gravitating object, there exists a vacuum and therefore its collapse does not cause to produce any gravitational waves. For this reason, one turns to a different fundamental symmetry, which is commonly known as cylindrical symmetry. By considering weak approximation fields, the solutions for gravitational waves produced by cylindrical stars have been constructed by Einstein and Rosen \cite{einstein1937gravitational}, and they also stated that such problems reduce to standard cylindrical waves in the case of Minkowski space. Many astrophysical facets have been investigated in relation to cylindrical symmetric propagation. Herrera and Santos \cite{herrera2005cylindrical} evaluated the smooth conditions to match the inner and outer geometries for gravitational collapse in cylindrical structure and sustained that the radial pressure at the boundary did not meet zero value. However, Herrera \emph{et al.} \cite{herrera2006matching} later noticed some errors in lengthy calculations, reporting that there would be no radial pressure at the boundary if such mistakes were corrected. Recently, Herrera \cite{herrera2017gibbs,herrera2020landauer} elaborated the physical interpretation of congruences as well as the fluid distribution in the study of astrophysical objects.

For static cylindrical self-gravitating star, Herrera \emph{et al.}
\cite{herrera2005static} established the matching criteria with the
Levi-Civita vacuum spacetime  and checked the regularity conditions
for them. They evaluated equations of motion and furthermore
demonstrated that an appropriate collection of conformally flat
solutions would represent an incompressible fluid. Sharif and Bhatti
\cite{sharif2012structure} found some structure scalars for charged
cylinder which contained anisotropic pressure and heat flux, and
examined its fundamental characteristics through those scalars. The
field equations and certain dynamical equations were also
constructed by considering the charge effects. By considering
cylindrical matter distribution by Sharif and Yousaf
\cite{sharif2012expansion}, vacuum cavity within the fluid
distribution was examined while they were studying the expansion
free condition. Their discussion focused on solutions which contain
a thin shell, and also on solutions that fulfil the Darmois
conditions on the boundary. The structural instability of an
isotropic cylindrical spacetime was studied by Sharif and Bhatti
\cite{sharif2014stability}, in which they observed the effects of
electromagnetic field. Their conclusion was that the energy density,
pressure and electric charge are the components which influence the
cylinder's stability. Yousaf \emph{et al.} \cite{yousaf2020study}
calculated the effective scalars by splitting the Riemann tensor
orthogonally in $f(R,T,Q)$ theory, found complexity factors for
cylindrical and charged spherical systems and also discussed
structural properties for these systems.

We have adopted a definition of complexity given in Herrera
\cite{herrera2018new} and want to generalize it $f(R,T,Q)$ theory of
gravity with static cylindrically symmetric spacetime. The scheme of
this paper is mentioned below. For $f(R,T,Q)$ gravity theory, in the
next section, we suggest some new definitions and the field
equations are also evaluated. Then in Sec. 3, we orthogonally split
the Riemann tensor to get some scalar factors, one of which is
claimed to be a complexity factor. After this, we evaluate the
condition to vanish the complexity factor in Sec. 4, and also
present some mathematical models in the context of $f(R,T,Q)$
theory. Eventually, all modified results and their effects have been
concluded in Sec. 5.

\section{Basic Definitions and Some Other Equations}

The fluid considered in this case is static self-gravitating
cylindrically symmetric which is coupled with electric charge and
pressure anisotropy. We study the stellar object in $f(R,T,Q)$
theory via some equations of motion, evaluate the C-energy as well
as Tolman mass functions and the required matching conditions on the
boundary surface $\Sigma$.

\subsection{Modified Field Equations}

For $f(R,T,Q)$ theory \cite{odintsov2013f,ayuso2015consistency,baffou2016exploring}, the Einstein-Hilbert action becomes after modification as
\begin{equation}\label{f1}
S=\frac{1}{2}\int \sqrt{-g}\left[f(R,T,Q)+L_{m}+L_{EM}\right]d^{4}x,
\end{equation}
where $L_{m}$ denotes the matter Lagrangian and the electromagnetic Lagrangian $L_{EM}$ contains the influence of electric charge.

Corresponding to \eqref{f1}, the field equations in this case are
\begin{equation}\label{f2}
G_{\rho\nu}=8\pi T_{\rho\nu}^{(eff)},
\end{equation}
where $G_{\rho\nu}$ is an Einstein tensor and $T_{\rho\nu}^{(eff)}$ could represent the stress energy tensor for $f(R,T,Q)$ theory,
\begin{align}
\nonumber T_{\rho\nu}^{(eff)}&=\frac{1}{f_{R}-L_{m}f_{Q}}\left[\left(f_{T}+\frac{1}{2}Rf_{Q}+1\right)X_{\rho\nu}+\left\{\frac{R}{2}(\frac{f}{R}-f_{R})-L_{m}f_{T}
\right.\right.\\\nonumber
&-\left.\frac{1}{2}\nabla_{\mu}\nabla_{\beta}(f_{Q}X^{\mu\beta})\right\}g_{\rho\nu}-\frac{1}{2}\Box(f_{Q}X_{\rho\nu})
-(g_{\rho\nu}\Box-\nabla_{\rho}\nabla_{\nu})f_{R}\\\label{f3}
&-2f_{Q}R_{\mu(\rho}X_{\nu)}^{\mu}+\nabla_{\mu}\nabla_{(\rho}[X_{\nu)}^{\mu}f_{Q}]+2(f_{Q}R^{\mu\beta}
+\left.f_{T}g^{\mu\beta})\frac{\partial^2 L_{m}}{\partial g^{\rho\nu}\partial g^{\mu\beta}}\right],
\end{align}
where $X_{\rho\nu}=T_{\rho\nu}^{(m)}+S_{\rho\nu}$. Also, $\nabla_{\rho}$ is the usual covariant derivative,~$R_{\rho\nu}$ is the Ricci tensor,~$g_{\rho\nu}$ is the metric tensor and $\Box\equiv g^{\gamma\alpha}\nabla_{\gamma}\nabla_{\alpha}$. Here, the subscripts $R,~T$ and $Q$ represent that their arguments are partially differentiated. The trace of usual stress energy tensor in GR provides a particular relation between $R$ and $T$. To study the charged fluid, the electromagnetic energy tensor $S_{\rho\nu}$ can be given as
\begin{equation}\label{f4}
  S_{\rho\nu}=\frac{1}{4\pi}\left(-F^{\gamma}_{\rho}F_{\nu\gamma}+\frac{1}{4}F^{\gamma\beta}F_{\gamma\beta}g_{\rho\nu}\right),
\end{equation}
where $F_{\rho\nu}=\phi_{\nu,\rho}-\phi_{\rho,\nu}$ is the Maxwell field tensor and in this case of static fluid, the four potential $\phi_{\rho}$ becomes as $\phi_{\rho}=\phi(r)\delta^{0}_{\rho}$. The electric charge becomes interior to radius $r$ as
\begin{equation}\label{f5}
  s(r)=4\pi \int_{0}^{r}\tau r^2Ydr,
\end{equation}
where $\tau$ denotes the charge density which depends on $r$ only. In this case, the matter Lagrangian shown in action \eqref{f1} is defined as $L_{m}=-\mu-\frac{1}{4}F^{\rho\nu}F_{\rho\nu}$, and after using the value of Maxwell tensor it becomes, $L_{m}=-\mu+\frac{s^2}{2r^4}$ in which $\mu$ is the energy density of the fluid.

Also, the usual energy-momentum tensor considered in \eqref{f3} is of the form
\begin{equation}\label{f6}
T_{\rho\nu}^{(m)}=\mu v_{\rho} v_{\nu}-Ph_{\rho\nu}+\Pi_{\rho\nu},
\end{equation}
where
\begin{equation}\label{f7}
\Pi_{\rho\nu}=\Pi\left(w_{\rho} w_{\nu}+\frac{1}{3}h_{\rho\nu}\right);\quad P=\frac{P_{r}+2P_{t}}{3},
\end{equation}
\begin{equation}\label{f8}
\Pi=P_{r}-P_{t};\quad h_{\rho\nu}=g_{\rho\nu}-v_\rho v_{\nu},
\end{equation}
and the 4-vector in radial direction $w^{\rho}$ is set as
\begin{equation}\label{f9}
w^\rho=(0,Y^{-1},0,0),
\end{equation}
with some properties $w^\mu v_{\mu}=0, w^\mu w_{\mu}=-1$.\\
Here, the term $\Pi$ represents anisotropic pressure which further contains the radial pressure $P_{r}$ and tangential pressure $P_{t}$. It should be noted that the pressure normally exists in three different directions for cylindrical anisotropic structure, but the pressure anisotropy considered in this case does not have the most general form.\\
Indeed, we found the following from Eq.\eqref{f3} as
\begin{align}\nonumber
&3\Box
f_R+\frac{1}{2}\Box(f_QX)-X(f_T+1)+\nabla_\pi\nabla_\rho(f_QX^{\pi\rho})+R(f_R-\frac{X}{2}f_Q)\\\nonumber
&+(Rf_Q+4f_T)\textit{L}_m-2f+2R_{\pi\rho}X^{\pi\rho}f_Q -2g^{\lambda\sigma}\frac{\partial^2\textit{L}_m}{\partial g^{\lambda\sigma}\partial g^{\pi\rho}}\left(f_Tg^{\pi\rho}+f_QR^{\pi\rho}\right)=0.
\end{align}
From above expression, the effects of $f(R,T)$ theory can be investigated with the assumption $Q=0$, while taking the vacuum case will help us to examine these complex results in $f(R)$ theory. Various authors \cite{odintsov2013f,ayuso2015consistency,baffou2016exploring} looked into the accurate analysis of their derivation and physical importance by examining our cosmic composition.

The 4-velocity vector in the comoving frame becomes
\begin{equation}\label{f10}
v^\rho=(X^{-1},0,0,0).
\end{equation}
The 4-acceleration $a^{\beta}=v^{\beta}_{;\pi}v^{\pi}$ gives only one non-disappearing component as follows
\begin{equation}\label{f11}
  a_{1}=-\frac{X'}{X}.
\end{equation}

In order to clarify the complexity, the appropriate inner space-time restricted by the hypersurface $\Sigma$ is indicated by a line element as
\begin{equation}\label{f12}
ds^2=-Y^2 dr^2-r^2(d\theta^2+\alpha^2 dz^2)+X^2 dt^2,
\end{equation}
where $X=X(r)$ and $Y=Y(r)$. Note that the above spacetime is not in
the most general form as it contains two independent functions, and
due to this assumption, the expressions for anisotropic tensor,
electric part of Weyl tensor and the scalar $Y_{TF}$ will be in
restricted forms. The most general form of the cylindrically
symmetric spacetime is given in \cite{Herrera2012}. For cylindrical
symmetry, we implement the following ranges on the above coordinates
as
\begin{equation}\nonumber
0\leq r, \quad 0\leq\theta\leq 2\pi, \quad -\infty< z< +\infty, \quad -\infty\leq t\leq +\infty.
\end{equation}

The cylindrical object \eqref{f12} delivers the field equations for $f(R,T,Q)$ theory as
\begin{align}\label{f13}
\frac{8\pi}{(f_{R}-L_{m}f_{Q})}\left[\mu^{(eff)}+Q_{0}^{(eff)}\right]&=-\frac{2Y'}{rY^3}+\frac{1}{r^2Y^2},\\\label{f14}
\frac{8\pi}{(f_{R}-L_{m}f_{Q})}\left[P_{r}^{(eff)}+Q_{1}^{(eff)}\right]&=-\frac{1}{r^2Y^2}-\frac{2X'}{rXY^2},\\\label{f15}
\frac{1}{(f_{R}-L_{m}f_{Q})}\left[P_{t}^{(eff)}+Q_{2}^{(eff)}\right]&=-\frac{X''}{XY^2}+\frac{Y'}{rY^3}+\frac{X'Y'}{XY^3}-\frac{X'}{rXY^2},
\end{align}
where $P_{r}^{(eff)}$,~$P_{t}^{(eff)}$ and $\mu^{(eff)}$ comprise the dark source terms of $f(R,T,Q)$ theory together with physical variables, while $Q_{1}^{(eff)}$,~$Q_{2}^{(eff)}$ and $Q_{0}^{(eff)}$ represent effective charge terms respectively. The values of all these quantities are set in Appendix A. Here, differentiation with respect to $r$ is shown by prime.

It is useful to say that the divergence of stress-energy tensor in $f(R,T,Q)$ theory is not conserved and therefore allowing all physical laws to be violated. Hence, the particles move in a non-geodesic path, owing to an extra force in its gravitational field. This can therefore be formalized as
\begin{align}\nonumber
\nabla^\lambda
X_{\lambda\epsilon}&=\frac{2}{Rf_Q+2f_T+1}\left[\nabla_\epsilon(\textit{L}_mf_T)
+\nabla_\lambda(f_QR^{\pi\lambda}X_{\pi\epsilon})-\frac{1}{2}(f_Tg_{\pi\rho}+f_QR_{\pi\rho})\right.\\\label{f16}
&\times\left.\nabla_\epsilon X^{\pi\rho}-G_{\lambda\epsilon}\nabla^\lambda(f_Q\textit{L}_m)\right].
\end{align}
From \eqref{f16}, the equation for the hydrostatic equilibrium can be read as
\begin{eqnarray}\nonumber
\left(\frac{P_{r}^{(eff)}+Q_{1}^{(eff)}}{H}\right)'&=&\frac{-X'}{HX}\left(\mu^{(eff)}+Q_{0}^{(eff)}+P^{(eff)}_{r}
+Q_{1}^{(eff)}\right)+\frac{2}{rH}\\\label{f17}
&\times&\left(P^{(eff)}_{t}+Q_{2}^{(eff)}-P^{(eff)}_{r}-Q_{1}^{(eff)}\right)+ZY^2,
\end{eqnarray}
where $H=f_{R}-L_{m}f_{Q}$. The term $Z$ represents extra curvature terms and refers to the constraints on the stress-energy tensor in $f(R,T,Q)$ theory whose value is given in Appendix A. It can be called as the generalized Tolman-Opphenheimer-Volkoff (TOV) equation which uses widely to study the compact stellar properties. In 1971, Bekenstein initially extended the TOV equation in GR for compact objects having the influence of electromagnetic field.

One can find the value of $\frac{X'}{X}$ from Eq.\eqref{f14} as
\begin{equation}\label{f18}
\frac{X'}{X}=\frac{4r^2}{r^2-8mr+4s^2}\left[-\frac{4\pi r}{H}\left(P_{r}^{(eff)}+Q_{1}^{(eff)}\right)+\frac{m}{r^2}-\frac{1}{8r}-\frac{s^2}{2r^3}\right].
\end{equation}
We will get the following by putting back Eq.\eqref{f18} in Eq.\eqref{f17},
\begin{align}\nonumber
\left(\frac{P_{r}^{(eff)}+Q_{1}^{(eff)}}{H}\right)'&=\frac{4r^2}{r^2-8mr+4s^2}\left[\frac{4\pi r}{H}\left(P_{r}^{(eff)}+Q_{1}^{(eff)}\right)-\frac{m}{r^2}\right.\\\nonumber
&+\left.\frac{1}{8r}+\frac{s^2}{2 r^3}\right]\left(\mu^{(eff)}+Q_{0}^{(eff)}+P^{(eff)}_{r}+Q_{1}^{(eff)}\right)\\\label{f19}
&+\frac{2}{rH}\left(P^{(eff)}_{t}+Q_{2}^{(eff)}-P^{(eff)}_{r}-Q_{1}^{(eff)}\right)+ZY^2,
\end{align}
in which the mass $m$ of the geometry is obtained as
\cite{PhysRev.136.B571}
\begin{equation}\label{f20}
m(r)\equiv
\tilde{E}=E\hat{l}=-\frac{r}{2Y^2}\left(1-\frac{Y^2}{4}\right)+\frac{s^2}{2r}.
\end{equation}
Equation \eqref{f20} can be written by combining it with
Eq.\eqref{f13} as
\begin{equation}\label{f21}
m(r)=\frac{r}{8}-4\pi\int_{0}^{r} \frac{\tilde{r}^2}{H}\left(\mu^{(eff)}+Q_{0}^{(eff)}\right) d\tilde{r}+\frac{s^2}{2r}.
\end{equation}

To study the cylindrical star, the line element which represents the exterior space-time is given as
\begin{equation}\label{f22}
ds^2=-\left(-\frac{2M(\nu)}{r}+\frac{S^2(\nu)}{r^2}\right)d\nu^2-2d\nu dr
+r^2(d\theta^2+\alpha^2 dz^2),
\end{equation}
where $M$ denotes its total mass. By succeeding the work done by Senovilla \cite{senovilla2013junction}, Yousaf \emph{et al.} \cite{Yousaf2017} introduced the corresponding criteria for smoothly matching of the space-times \eqref{f12} and \eqref{f22} in $f(R,T,Q)$ theory. Therefore, at the hypersurface $r=r_{\Sigma}=$constant, we impose some constraints on the effective radial pressure and its charge component separately as
\begin{equation}\label{f23}
\tilde{E}-M_=^{\Sigma}\frac{1}{8}; \quad [P_{r}]_=^{\Sigma}-F_{0}; \quad s^2{_=^{\Sigma}}\tilde{H}S^2-8\pi r^4F_{1}.
\end{equation}
The values of $F_{0}$ and $F_{1}$ are set out in Appendix A. Note that the conditions \eqref{f23} are necessarily required to match the inner and outer space-times on $\Sigma$.

\subsection{Some Curvature Tensors}

A well known tensor which is used to measure the space-time curvature named as the Riemann tensor can be written as a linear combination of the Weyl tensor $C_{\alpha\gamma\beta\rho}$, the Ricci tensor $R_{\gamma\rho}$ and the Ricci scalar $R$ as
\begin{eqnarray}\nonumber
R^\alpha_{\pi\beta\rho}&=&C^\alpha_{\pi\beta\rho}-\frac{1}{2}R_{\pi\beta}\delta^\alpha_{\rho}+\frac{1}{2}R_{\pi\rho}\delta^\alpha_{\beta}
-\frac{1}{2}R^\alpha_{\rho}g_{\pi\beta}+\frac{1}{2}R^\alpha_{\beta}g_{\pi\rho}\\\label{f24}
&-&\frac{1}{6}R\left(\delta^\alpha_{\beta}g_{\pi\rho}-g_{\pi\beta}\delta^\alpha_{\rho}\right).
\end{eqnarray}
One can define the electric and magnetic part of Weyl tensor separately as
\begin{equation}\label{f25}
E_{\gamma\beta}=C_{\gamma\delta\beta\pi}v^{\delta}v^{\pi}, \quad H_{\gamma\beta}=\frac{1}{2}\eta_{\gamma\delta\mu\pi}C^{\mu\pi}_{\quad\beta\sigma}v^{\delta}v^{\sigma}.
\end{equation}
The magnetic part of Weyl tensor described in Eq.\eqref{f25} vanishes, as we have considered static cylindrical structure \eqref{f12}. Thus another form of the Weyl tensor may be written as
\begin{equation}\label{f26}
C_{\pi\lambda\kappa\rho}=(g_{\pi\lambda\mu\nu}g_{\kappa\rho\gamma\beta}
-\eta_{\pi\lambda\mu\nu}\eta_{\kappa\rho\gamma\beta})v^\mu v^\gamma E^{\nu\beta},
\end{equation}
where $g_{\pi\lambda\alpha\nu}=g_{\pi\alpha}g_{\lambda\nu}-g_{\pi\nu}g_{\lambda\alpha}$, and $\eta_{\lambda\pi\alpha\nu}$ is the Levi-Civita tensor. The electric part $E_{\rho\nu}$ can further be written as
\begin{equation}\label{f27}
E_{\rho\nu}=E\left(w_{\rho}w_{\nu}+\frac{1}{3}h_{\rho\nu}\right),
\end{equation}
with its non-fading elements
\begin{equation}\label{f28}
E_{11}=\frac{2}{3}Y^2E, \quad E_{22}=-\frac{1}{3}r^2E,
\end{equation}
where
\begin{equation}\label{f29}
E=\frac{1}{2XY^2}\left[X''-\frac{X'Y'}{Y}+\frac{XY'}{rY}-\frac{X'}{r}+\frac{X}{r^2}\right],
\end{equation}
which must meet some conditions given below as
\begin{equation}\label{f30}
E^\rho_{\rho}=0,\quad E_{\rho\gamma}=E_{(\rho\gamma)},\quad E_{\rho\gamma}v^\gamma=0.
\end{equation}

\subsection{The Mass Function}

Here, we will evaluate few useful equations by following the definitions of C-energy \cite{thorne1965energy} and Tolman mass \cite{tolman1930use} in order to investigate some properties of cylindrical system. Such equations will further support us to make the relation between mass function and Conformal tensor. Equations \eqref{f2},~\eqref{f20},~\eqref{f24} and \eqref{f27} are used to write the following as
\begin{align}\nonumber
m&=\frac{r}{8}-\frac{4\pi r^3}{3H}\left(\mu^{(eff)}+Q^{(eff)}_{0}+P^{(eff)}_{t}+Q^{(eff)}_{2}
-P^{(eff)}_{r}-Q^{(eff)}_{1}\right)\\\label{f31}
&+\frac{s^2}{2r}-\frac{r^3E}{3}.
\end{align}
One may also write above expression for $E$ as
\begin{eqnarray}\nonumber
E&=&-\frac{4\pi}{r^3}\int_{0}^{r} \tilde{r}^3\left(\frac{\mu^{(eff)}+Q^{(eff)}_{0}}{H}\right)' d\tilde{r}+\frac{4\pi}{H} \left(P^{(eff)}_{r}+Q^{(eff)}_{1}\right.\\\label{f32}
&-&\left.P^{(eff)}_{t}-Q^{(eff)}_{2}\right).
\end{eqnarray}
Note that Eq.\eqref{f32} presents the relation of Weyl tensor with cylindrical structural properties, i.e., effective energy density inhomogeneity, effective local anisotropic pressure and electric charge of the fluid with modified corrections. One can insert Eq.\eqref{f32} in Eq.\eqref{f31} and get
\begin{equation}\label{f33}
m(r)=\frac{r}{8}+\frac{s^2}{2r}-\frac{4\pi r^3}{3H}\left(\mu^{(eff)}+Q^{(eff)}_{0}\right)+\frac{4\pi}{3}\int_{0}^{r} \tilde{r}^3 \left(\frac{\mu^{(eff)}+Q^{(eff)}_{0}}{H}\right)' d\tilde{r},
\end{equation}
which contributes to link the mass of our geometry with uniform energy density and charge distribution. The formula mentioned above, allows us to analyze the changes in mass function which are carried by effective inhomogeneous energy density and respective component of effective charge for self-gravitating system.

Another way in order to measure the total energy of static cylinder was suggested by Tolman \cite{tolman1930use} which is given as
\begin{equation}\label{f34}
m_{T}=4\pi \int_{0}^{r_{\Sigma}} r^2XY\left(T_{0}^{0(eff)}-T_{1}^{1(eff)}-2T_{2}^{2(eff)}\right)dr.
\end{equation}
With and without taking into consideration the effects of electric charge in self-gravitating structures, Bhatti \emph{et al.} \cite{zaeem2019energy,bhatti2019tolman} evaluated the Tolman mass function in $f(R)$ theory. Thus, within the cylindrical configuration of radius $r$, Eq.\eqref{f34} becomes
\begin{equation}\label{f35}
m_{T}=4\pi \int_{0}^{r} r^2XY\left(T_{0}^{0(eff)}-T_{1}^{1(eff)}-2T_{2}^{2(eff)}\right)dr.
\end{equation}
By using field equations in Eq.\eqref{f35}, we receive
\begin{equation}\label{f36}
m_{T}=-\frac{X'r^2}{Y}-4\pi \int_{0}^{r}\frac{XYr^2Q_{3}^{(eff)}}{H}dr,
\end{equation}
where $Q_{3}^{(eff)}=Q_{0}^{(eff)}+Q_{1}^{(eff)}+2Q_{2}^{(eff)}$, or it becomes by putting the value of $X'$ from Eq.\eqref{f18} as
\begin{equation}\label{f37}
m_{T}=XY\left[\frac{4\pi r^3}{H}\left(P_{r}^{(eff)}+Q_{1}^{(eff)}\right)-m+\frac{s^2}{2r}+\frac{r}{8}\right].
\end{equation}
This expression could be referred to as the effective gravitational mass. The gravitational acceleration $(a=-w^{\nu}a_{\nu})$ of a test molecule can be pursued instantaneously at rest in a static gravitational field as
\begin{equation}\label{f38}
a=\frac{X'}{XY}=-\frac{1}{Xr^2}\left[m_{T}+4\pi\int_{0}^{r}\frac{XYr^2Q_{3}^{(eff)}}{H}dr\right].
\end{equation}
Equation \eqref{f36} can be more appropriately expressed as
\begin{eqnarray}\nonumber
m_{T}&=&(m_{T})_{\Sigma}(\frac{r}{r_{\Sigma}})^3+r^3\int_{r}^{r_{\Sigma}}\frac{XY}{\tilde{r}}\left[\frac{4\pi}{H}\Pi^{(eff)}+E\right]d\tilde{r}\\\label{f39}
&+&4\pi r^3 \int_{r}^{r_{\Sigma}}\frac{XYQ_{4}^{(eff)}}{\tilde{r}H}d\tilde{r},
\end{eqnarray}
where $Q_{4}^{(eff)}=Q_{0}^{(eff)}+2Q_{1}^{(eff)}+Q_{2}^{(eff)}$. One can get from the above expression by using Eq.\eqref{f32} in it, as
\begin{eqnarray}\nonumber
m_{T}&=&(m_{T})_{\Sigma}(\frac{r}{r_{\Sigma}})^3+r^3\int_{r}^{r_{\Sigma}}\frac{XY}{\tilde{r}}\left[\frac{8\pi}{H}\Pi^{(eff)}
-\frac{4\pi}{\tilde{r}^3}\int_{0}^{r}\tilde{r^3}\left(\frac{\mu^{(eff)}}{H}\right)'d\tilde{r}\right]d\tilde{r}\\\label{f40}
&+&4\pi r^3 \int_{r}^{r_{\Sigma}}\frac{XYQ_{5}^{(eff)}}{\tilde{r}H}d\tilde{r},
\end{eqnarray}
where $Q_{5}^{(eff)}=Q_{0}^{(eff)}+3Q_{1}^{(eff)}$. Equation \eqref{f40} could be useful to interpret the relation between effective Tolman mass and some physical phenomenon like effective inhomogeneity of the energy density, effective local pressure anisotropy and effective form of electric charge in the static cylinder.

\section{The Orthogonal Splitting of The Riemann Tensor}

The orthogonal splitting of the Riemann tensor was first suggested by Bel \cite{bel1961inductions}
and Herrera \emph{et al.} \cite{herrera2004spherically} from which one can obtain the following three tensors as
\begin{eqnarray}\label{f41}
&&Y_{\delta\gamma}=R_{\delta\beta\gamma\alpha}u^\beta u^\alpha,\\\label{f42}
&&Z_{\delta\gamma}=*R_{\delta\beta\gamma\alpha}u^\beta u^\alpha=\frac{1}{2}\eta_{\delta\beta\epsilon\mu}R^{\epsilon\mu}_{\gamma\alpha}u^\beta u^\alpha,\\\label{f43}
&&X_{\delta\gamma}=*R^*_{\delta\beta\gamma\alpha}u^\beta u^\alpha=\frac{1}{2}\eta_{\delta\beta}^{\epsilon\mu}R^*_{\epsilon\mu\gamma\alpha}u^\beta u^\alpha,
\end{eqnarray}
where $*$ is the dual tensor and $\eta_{\delta\epsilon\mu\beta}$ is
known as Levi-Civita symbol which takes $0$ value for no
permutation,~$1$ for even permutation and $-1$ for odd permutaion.
Now we use field equations to rewrite the Riemann tensor \eqref{f24}
by means of above tensors \eqref{f41}-\eqref{f43} (see
\cite{gomez2007dynamical}) as
\begin{equation}\label{f44}
R^{\delta\beta}_{\quad\gamma\alpha}=C^{\delta\beta}_{\quad\gamma\alpha}+16\pi T^{(eff)[\delta}_{[\gamma} \delta_{\alpha]}^{\beta]}+8\pi T^{(eff)}\left(\frac{1}{3}\delta^{\delta}_{\quad[\gamma} \delta_{\alpha]}^{\beta}-\delta^{[\delta}_{\quad[\gamma} \delta_{\alpha]}^{\beta]}\right).
\end{equation}
The Riemann tensor \eqref{f44} can further be divided into three components after using Eq.\eqref{f3} in it as
\begin{equation}\label{f45}
R^{\delta\beta}_{\quad\gamma\alpha}=R^{\delta\beta}_{(I)\gamma\alpha}+R^{\delta\beta}_{(II)\gamma\alpha}+R^{\delta\beta}_{(III)\gamma\alpha},
\end{equation}
where
\begin{align}\nonumber
R^{\delta\beta}_{(I)\gamma\alpha}&=\frac{16\pi}{H}(f_{T}+\frac{1}{2}Rf_{Q}+1) \left[\left(\mu+\frac{s^2}{8\pi r^4}\right) v^{[\delta}v_{[\gamma} \delta_{\alpha]}^{\beta]}-\left(P+\frac{s^2}{24\pi r^4}\right)\right.\\\nonumber
&\times\left.h^{[\delta}_{\quad[\gamma} \delta_{\alpha]}^{\beta]}+\Pi^{[\delta}_{\quad[\gamma} \delta_{\alpha]}^{\beta]}-\frac{s^2}{4\pi r^4}\left(w^{[\delta}w_{[\gamma} \delta_{\alpha]}^{\beta]}+\frac{1}{3}h^{[\delta}_{\quad[\gamma} \delta_{\alpha]}^{\beta]}\right)\right]+\frac{8\pi}{H}\\\nonumber
&\times\left[(f_{T}+\frac{1}{2}Rf_{Q}+1)(\mu-3P)+4\left\{\frac{R}{2}\left(\frac{f}{R}
-f_{R}\right)+\left(\mu-\frac{s^2}{2r^4}\right)f_{T}\right.\right.\\\nonumber
&-\left.\frac{1}{2}\nabla_{\mu}\nabla_{\nu}(f_{Q}X^{\mu\nu})\right\}-3\Box f_{R}-\frac{1}{2}\Box\{f_{Q}(\mu-3P)\}+\nabla_{\mu}\nabla_{\pi}(f_{Q}X^{\mu\pi})\\\nonumber
&-\left.2f_{Q}R_{\mu\pi}X^{\mu\pi}+2g^{\pi\xi}(f_{Q}R^{\mu\nu}+f_{T}g^{\mu\nu})\frac{\partial^2L_{m}}{\partial g^{\pi\xi}\partial g^{\mu\nu}}\right]\\\label{f46}
&\times\left(\frac{1}{3}\delta^{\delta}_{\quad[\gamma} \delta_{\alpha]}^{\beta}-\delta^{[\delta}_{\quad[\gamma} \delta_{\alpha]}^{\beta]}\right),
\\\nonumber
R^{\delta\beta}_{(II)\gamma\alpha}&=\frac{4\pi}{H}\left[2\left\{\frac{R}{2}
\left(\frac{f}{R}-f_{R}\right)+\left(\mu-\frac{s^2}{2r^4}\right) f_{T}-\frac{1}{2}\nabla_{\mu}\nabla_{\rho}(f_{Q}X^{\mu\rho})\right\}\right.\\\nonumber
&\times\left(\delta^{\delta}_{\gamma} \delta_{\alpha}^{\beta}-\delta^{\delta}_{\alpha} \delta_{\gamma}^{\beta}\right)-\frac{1}{2}\Box\left\{f_{Q}\left(X^{\delta}_{\gamma} \delta_{\alpha}^{\beta}-X^{\delta}_{\alpha} \delta_{\gamma}^{\beta}-X^{\beta}_{\gamma} \delta_{\alpha}^{\delta}+X^{\beta}_{\alpha} \delta_{\gamma}^{\delta}\right)\right\}\\\nonumber
&-2\Box f_{R}\left(\delta^{\delta}_{\gamma} \delta_{\alpha}^{\beta}-\delta^{\delta}_{\alpha} \delta_{\gamma}^{\beta}\right)+\left( \delta_{\alpha}^{\beta}\nabla^{\delta}\nabla_{\gamma}- \delta_{\gamma}^{\beta}\nabla^{\delta}\nabla_{\alpha}- \delta_{\alpha}^{\delta}\nabla^{\beta}\nabla_{\gamma}\right.\\\nonumber
&+\left. \delta_{\gamma}^{\delta}\nabla^{\beta}\nabla_{\alpha}\right)f_{R}
-f_{Q}\left(R_{\mu\gamma}X^{\mu\delta} \delta_{\alpha}^{\beta}-R_{\mu\alpha}X^{\mu\delta} \delta_{\gamma}^{\beta}
-R_{\mu\gamma}X^{\mu\beta} \delta_{\alpha}^{\delta}\right.\\\nonumber
&+\left.R_{\mu\alpha}X^{\mu\beta} \delta_{\gamma}^{\delta}+R^{\delta}_{\mu}X^{\mu}_{\gamma} \delta_{\alpha}^{\beta}-R^{\delta}_{\mu}X^{\mu}_{\alpha} \delta_{\gamma}^{\beta}-R^{\beta}_{\mu}X^{\mu}_{\gamma} \delta_{\alpha}^{\delta}+R^{\beta}_{\mu}X^{\mu}_{\alpha} \delta_{\gamma}^{\delta}\right)\\\nonumber
&+\frac{1}{2}\nabla_{\mu}\nabla^{\delta}\left\{f_{Q}\left(X^{\mu}_{\gamma} \delta_{\alpha}^{\beta}-X^{\mu}_{\alpha} \delta_{\gamma}^{\beta}\right)\right\}+\frac{1}{2}\nabla_{\mu}\nabla_{\alpha}\left\{f_{Q}\left(X^{\beta\mu} \delta_{\gamma}^{\delta}-X^{\delta\mu} \delta_{\gamma}^{\beta}\right)\right\}\\\nonumber
&+\frac{1}{2}\nabla_{\mu}\nabla^{\beta}\left\{f_{Q}\left(X^{\mu}_{\alpha} \delta_{\gamma}^{\delta}-X^{\mu}_{\gamma} \delta_{\alpha}^{\delta}\right)\right\}+\frac{1}{2}\nabla_{\mu}\nabla_{\gamma}\left\{f_{Q}\left(X^{\delta\mu} \delta_{\alpha}^{\beta}-X^{\beta\mu} \delta_{\alpha}^{\delta}\right)\right\}\\\nonumber
&+2g^{\delta\beta}(f_{Q}R^{\mu\rho}+f_{T}g^{\mu\rho})\left\{\delta_{\alpha}^{\pi}\frac{\partial^2L_{m}}{\partial g^{\pi\gamma}\partial g^{\mu\rho}}-\delta_{\gamma}^{\pi}\frac{\partial^2L_{m}}{\partial g^{\pi\alpha}\partial g^{\mu\rho}}\right.\\\label{f47}
&-\delta_{\alpha}^{\pi}\frac{\partial^2L_{m}}{\partial g^{\pi\gamma}\partial g^{\mu\rho}}+\left.\left.\delta_{\gamma}^{\pi}\frac{\partial^2L_{m}}{\partial g^{\pi\alpha}\partial g^{\mu\rho}}\right\}\right],
\\\label{f48}
R^{\delta\beta}_{(III)\gamma\alpha}&=4v^{[\delta}v_{[\gamma} E_{\alpha]}^{\beta]}-\epsilon^{\delta\beta}_{\mu}\epsilon_{\gamma\alpha\rho}E^{\mu\rho},
\end{align}
with
\begin{equation}\label{f49}
\epsilon_{\mu\beta\rho}=v^\delta \eta_{\delta\mu\beta\rho},\quad \epsilon_{\mu\beta\gamma}v^\gamma=0,
\end{equation}
where the magnetic part of Weyl tensor has been disappeared for the considered static cylindrical symmetry.

It is possible to find the tensors $X_{\delta\gamma},~Y_{\delta\gamma}$ and $Z_{\delta\gamma}$ which encompass material variables, electric charge and extra curvature terms of $f(R,T,Q)$ gravity, as
\begin{align}\nonumber
Y_{\delta\gamma}&=E_{\delta\gamma}+\frac{1}{H}\left\{\frac{4\pi}{3}(\mu+3P)h_{\delta\gamma}+4\pi \Pi_{\delta\gamma}-\frac{s^2}{r^4}\left(w_{\delta}w_{\gamma}+\frac{1}{3}h_{\delta\gamma}\right)+\frac{s^2}{3r^4}h_{\delta\gamma}\right\}\\\nonumber
&\times(f_{T}+\frac{1}{2}Rf_{Q}+1)-\frac{8\pi}{3H}\left\{\frac{R}{2}\left(\frac{f}{R}-f_{R}\right)+\left(\mu-\frac{s^2}{2r^4}\right) f_{T}\right.\\\nonumber
&-\left.\frac{1}{2}\nabla_{\mu}\nabla_{\rho}(f_{Q}X^{\mu\rho})\right\}h_{\delta\gamma}+\frac{4\pi}{H}\left[-\frac{1}{2}\left\{\Box(f_{Q}X_{\delta\gamma})-v_{\delta}v_{\beta}\Box(f_{Q}X^{\beta}_{\gamma})\right.\right.\\\nonumber
&-\left.v_{\gamma}v^{\alpha}\Box(f_{Q}X_{\delta\alpha})
+g_{\delta\gamma}v_{\beta}v^{\alpha}\Box(f_{Q}X^{\beta}_{\alpha})\right\}
+(\nabla_{\delta}\nabla_{\gamma}f_{R}-v_{\delta}v_{\beta}\nabla^{\beta}\nabla_{\gamma}f_{R}\\\nonumber
&-v_{\gamma}v^{\alpha}\nabla_{\delta}\nabla_{\alpha}f_{R}
+g_{\delta\gamma}v_{\beta}v^{\alpha}\nabla^{\beta}\nabla_{\alpha}f_{R})
+2f_{Q}R_{\delta\mu}\left\{\left(P+\frac{s^2}{24\pi r^4}\right) h^{\mu}_{\gamma}\right.\\\nonumber
&-\left.\left(w^{\mu}w_{\gamma}+\frac{1}{3}h^{\mu}_{\gamma}\right)\left(\Pi-\frac{s^2}{4\pi r^4}\right)\right\}
-f_{Q}R^{\beta}_{\mu}\left\{\left(\mu+\frac{s^2}{8\pi r^4}\right) v^{\mu}v_{\beta}h_{\delta\gamma}\right.\\\nonumber
&\left.-v_{\delta}v_{\beta}\left(w^{\mu}w_{\gamma}+\frac{1}{3}h^{\mu}_{\gamma}\right)\left(\Pi-\frac{s^2}{4\pi r^4}\right)+v_{\delta}v_{\beta}\left(P+\frac{s^2}{24\pi r^4}\right) h^{\mu}_{\gamma}\right\}\\\nonumber
&-f_{Q}R_{\mu\alpha}\left\{\left(\mu+\frac{s^2}{8\pi r^4}\right) v^{\mu}v^{\alpha}h_{\delta\gamma}-v_{\gamma}v^{\alpha}\left(w^{\mu}w_{\delta}+\frac{1}{3}h^{\mu}_{\delta}\right)\left(\Pi-\frac{s^2}{4\pi r^4}\right)\right.\\\nonumber
&+\left.v_{\gamma}v^{\alpha}\left(P+\frac{s^2}{24\pi r^4}\right)h^{\mu}_{\delta}\right\}+\frac{1}{2}\{\nabla_{\mu}\nabla_{\delta}(f_{Q}X^{\mu}_{\gamma})
+g_{\delta\gamma}v_{\beta}v^{\alpha}\nabla_{\mu}\nabla^{\beta}(f_{Q}X^{\mu}_{\alpha})\\\nonumber
&+\nabla_{\mu}\nabla_{\gamma}(f_{Q}X^{\mu}_{\delta})
+g_{\delta\gamma}v_{\beta}v^{\alpha}\nabla_{\mu}\nabla_{\alpha}(f_{Q}X^{\mu\beta})-v_{\gamma}v^{\alpha}\nabla_{\mu}\nabla_{\delta}(f_{Q}X^{\mu}_{\alpha})\\\nonumber
&-v_{\beta}v_{\delta}\nabla_{\mu}\nabla_{\gamma}(f_{Q}X^{\beta\mu})-v_{\beta}v_{\delta}\nabla_{\mu}\nabla^{\beta}(f_{Q}X^{\mu}_{\gamma})
-v_{\gamma}v^{\alpha}\nabla_{\mu}\nabla_{\alpha}(f_{Q}X^{\mu}_{\delta})\}\\\nonumber
&+\left.2h^{\epsilon}_{\delta}(f_{Q}R^{\mu\rho}+f_{T}g^{\mu\rho})\frac{\partial^2L_{m}}{\partial g^{\epsilon\gamma}\partial g^{\mu\rho}}\right]+\frac{8\pi}{3H}\left[\frac{1}{2}\Box\{f_{Q}(\mu-3P)\}+2f_{Q}R_{\mu\epsilon}\right.\\\nonumber
&\times\left\{\left(\mu+\frac{s^2}{8\pi r^4}\right) v^{\mu}v^{\epsilon}
+\left(\Pi-\frac{s^2}{4\pi r^4}\right)\left(w^{\mu}w^{\epsilon}+\frac{1}{3}h^{\mu\epsilon}\right)-h^{\mu\epsilon}\right.\\\label{f50}
&\times\left.\left.\left(P+\frac{s^2}{24\pi r^4}\right)\right\}
-\nabla_{\mu}\nabla_{\epsilon}(f_{Q}X^{\mu\epsilon})-2g^{\epsilon\xi}(f_{Q}R^{\mu\rho}+f_{T}g^{\mu\rho})\frac{\partial^2L_{m}}{\partial g^{\epsilon\xi}\partial g^{\mu\rho}}\right]h_{\delta\gamma},
\\\nonumber
Z_{\delta\gamma}&=\frac{4\pi}{H}\left[\frac{1}{2}v^{\alpha}\Box(f_{Q}X^{\epsilon}_{\alpha})+f_{Q}\left(\mu+\frac{s^2}{8\pi r^4}\right) R_{\alpha}^{\epsilon}v^{\alpha}-v^{\alpha}\nabla^{\epsilon}\nabla_{\alpha}f_{R}\right.\\\nonumber
&+\frac{1}{3}f_{Q}\left(\Pi-\frac{s^2}{4\pi r^4}\right) R^{\epsilon}_{\alpha}v^{\alpha}-f_{Q}\left(P+\frac{s^2}{24\pi r^4}\right)R_{\alpha}^{\epsilon}v^{\alpha}-\frac{1}{2}v^{\alpha}\nabla_{\mu}\nabla_{\alpha}(f_{Q}X^{\mu\epsilon})\\\label{f51}
&-\left.\frac{1}{2}v^{\alpha}\nabla_{\mu}\nabla^{\epsilon}(f_{Q}X^{\mu}_{\alpha})\right]\epsilon_{\epsilon\gamma\delta},
\end{align}
and
\begin{align}\nonumber
X_{\delta\gamma}&=-E_{\delta\gamma}+\frac{1}{H}\left\{\frac{8\pi}{3}\mu h_{\delta\gamma}+4\pi \Pi_{\delta\gamma}-\frac{s^2}{r^4}\left(w_{\delta}w_{\gamma}+\frac{1}{3}h_{\delta\gamma}\right)+\frac{s^2}{3r^4}h_{\delta\gamma}\right\}\\\nonumber
&\times(f_{T}+\frac{1}{2}Rf_{Q}+1)+\frac{4\pi}{H}\left[\left\{\nabla^{\nu}\nabla_{\epsilon}f_{R}-\frac{1}{2}\Box(f_{Q}X^{\nu}_{\epsilon})
+\frac{1}{2}\nabla_{\mu}\nabla_{\epsilon}(f_{Q}X^{\mu\nu})\right.\right.\\\nonumber
&+\left.\left.\frac{1}{2}\nabla_{\mu}\nabla^{\nu}(f_{Q}X^{\mu}_{\epsilon})\right\}\epsilon^{\epsilon\alpha}_{\delta}\epsilon_{\nu\alpha\gamma}
+f_{Q}R_{\mu\epsilon}\left\{\left(P+\frac{s^2}{24\pi r^4}\right)-\frac{1}{3}\left(\Pi-\frac{s^2}{4\pi r^4}\right)\right\}\right.\\\nonumber
&\times\left.\epsilon^{\epsilon\alpha}_{\delta}\epsilon^{\mu}_{\alpha\gamma}+f_{Q}R^{\nu}_{\mu}\left\{\left(P+\frac{s^2}{24\pi r^4}\right)-\frac{1}{3}\left(\Pi-\frac{s^2}{4\pi r^4}\right)\right\}\epsilon^{\mu\alpha}_{\delta}\epsilon_{\nu\alpha\gamma}\right]\\\nonumber
&+\frac{8\pi}{3H}\left[\left\{\frac{R}{2}\left(\frac{f}{R}-f_{R}\right)+\left(\mu-\frac{s^2}{2r^4}\right) f_{T}-\frac{1}{2}\nabla_{\mu}\nabla_{\rho}(f_{Q}X^{\mu\rho})\right\}\right.\\\nonumber
&-\frac{1}{2}\Box
\{f_{Q}(\mu-3P)\}-2Rf_{Q}\left\{\frac{1}{3}\left(\Pi-\frac{s^2}{4\pi r^4}\right)-\left(P+\frac{s^2}{24\pi r^4}\right)\right\}\\\label{f52}
&+\left.\nabla_{\mu}\nabla_{\nu}(f_{Q}X^{\mu\nu})+2g^{\nu\epsilon}(f_{Q}R^{\mu\rho}+f_{T}g^{\mu\rho})\frac{\partial^2L_{m}}{\partial g^{\nu\epsilon}\partial g^{\mu\rho}}\right]h_{\delta\gamma}.
\end{align}
The three tensors \eqref{f50}-\eqref{f52} can further be expressed
with the help of following scalars $X_{T},~X_{TF},~Y_{T}$ and
$Y_{TF}$ whose values are given as
\begin{eqnarray}\label{f53}
X_{T}&=&\frac{1}{H}\left(8\pi\mu+\frac{s^2}{r^4}\right)\left(1+\frac{1}{2}Rf_{Q}+f_{T}\right)+\varphi_{1}^{(D)},
\\\label{f54}
X_{TF}&=&-E+\frac{1}{H}\left(4\pi\Pi-\frac{s^2}{r^4}\right)\left(1+\frac{1}{2}Rf_{Q}+f_{T}\right),
\\\label{f55}
Y_{T}&=&\frac{4\pi}{H}\left(\mu+3P_{r}-2\Pi+\frac{s^2}{4\pi r^4}\right)\left(1+\frac{1}{2}Rf_{Q}+f_{T}\right) +\varphi_{2}^{(D)},
\\\label{f56}
Y_{TF}&=&E+\frac{1}{H}\left(4\pi\Pi-\frac{s^2}{r^4}\right)\left(1+\frac{1}{2}Rf_{Q}+f_{T}\right)+\varphi_{3}^{(D)},
\end{eqnarray}
where
$\varphi_{3}^{(D)}=\frac{1}{w_{\delta}w_{\gamma}+\frac{1}{3}h_{\delta\gamma}}\varphi_{\delta\gamma}^{(D)}$.
These scalars were first determined and discussed in detail by
Herrera \emph{et al.} \cite{herrera2009structure}. Combining
Eqs.\eqref{f54} and \eqref{f56} with Eq.\eqref{f32} separately, we
have respectively
\begin{align}\nonumber
X_{TF}&=\frac{4\pi}{r^3}\int_{0}^{r}\tilde{r}^3\left(\frac{\mu^{(eff)}+Q_{0}^{(eff)}}{H}\right)'d\tilde{r}
-\frac{4\pi}{H}(\Pi^{(eff)}+Q_{1}^{(eff)}-Q_{2}^{(eff)})\\\label{f57}
&+\frac{1}{H}\left(4\pi\Pi-\frac{s^2}{r^4}\right)\left(1+\frac{1}{2}Rf_{Q}+f_{T}\right),
\\\nonumber
Y_{TF}&=-\frac{4\pi}{r^3}\int_{0}^{r}\tilde{r}^3\left(\frac{\mu^{(eff)}+Q_{0}^{(eff)}}{H}\right)'d\tilde{r}
+\frac{4\pi}{H}(\Pi^{(eff)}+Q_{1}^{(eff)}-Q_{2}^{(eff)})\\\label{f58}
&+\frac{1}{H}\left(4\pi\Pi-\frac{s^2}{r^4}\right)\left(1+\frac{1}{2}Rf_{Q}+f_{T}\right)+\varphi_{3}^{(D)},
\end{align}
whereas Appendix B contains the values of $\varphi_{1}^{(D)}$, $\varphi_{2}^{(D)}$ and $\varphi_{\delta\gamma}^{(D)}$.

By adding two above scalars $X_{TF}$ and $Y_{TF}$, we get local anisotropic pressure and electric charge along with corrections terms of $f(R,T,Q)$ theory, as
\begin{equation}\label{f59}
X_{TF}+Y_{TF}=\frac{2}{H}\left(4\pi\Pi-\frac{s^2}{r^4}\right)\left(1+\frac{1}{2}Rf_{Q}+f_{T}\right)+\varphi_{3}^{(D)}.
\end{equation}
The physical significance of the scalar function $Y_{TF}$ can be determined by utilizing Eq.\eqref{f56} in Eq.\eqref{f39} as
\begin{eqnarray}\nonumber
m_{T}&=&(m_{T})_{\Sigma}(\frac{r}{r_{\Sigma}})^3+r^3\int_{r}^{r_{\Sigma}}\frac{XY}{\tilde{r}}\left[Y_{TF}
+\frac{4\pi\Pi^{(eff)}}{H}-\frac{1}{H}\left(4\pi\Pi-\frac{s^2}{r^4}\right)\right.\\\label{f60}
&\times&\left.\left(1+\frac{1}{2}Rf_{Q}+f_{T}\right)-\varphi_{3}^{(D)}\right]d\tilde{r}+4\pi r^3\int_{r}^{r_{\Sigma}}\frac{XY}{\tilde{r}H}Q_{4}^{(eff)}d\tilde{r}.
\end{eqnarray}
One can observe by comparing Eq.\eqref{f39} with Eq.\eqref{f60} that the scalar $Y_{TF}$ contributes to determine the impact of anisotropic pressure, energy density inhomogeneity and electromagnetic field on the Tolman mass along with dark source terms of $f(R,T,Q)$ theory. Likewise, in terms of scalar factor $Y_{T}$, the Tolman mass \eqref{f35} takes the form as
\begin{eqnarray}\nonumber
m_{T}&=&\int_{0}^{r}XY\tilde{r}^2\left[Y_{T}-\frac{4\pi}{H}\left(\mu+3P_{r}-2\Pi+\frac{s^2}{4\pi r^4}\right)\left(\frac{1}{2}Rf_{Q}+f_{T}\right)\right.\\\label{f61}
&-&\left.\frac{4\pi s^2}{Hr^4}+\frac{4\pi}{H}\left(\mu^{(D)}+P_{r}^{(D)}+2P_{t}^{(D)}+Q_{3}^{(eff)}\right)-\varphi_{2}^{(D)}\right]d\tilde{r}.
\end{eqnarray}
Equation \eqref{f61} provides the relation among the scalar $Y_{T}$, matter variables and the Tolman mass which are influenced by modified correction terms. Herrera \emph{et al.} \cite{herrera2011meaning,Herrera2012} and Yousaf \emph{et al.} \cite{PhysRevD.95.024024,bhatti2017dynamical,bhatti2017evolution} observed the role of the scalar $Y_{T}$ on equation for evolution which is commonly called as Raychaudhuri equation, and thus, Eq.\eqref{f61} could be used to provide this equation through $m_{T}$ even for charged structure in $f(R,T,Q)$ theory.

\section{Matter Distribution With Disappearing Complexity Factor}

It should be obvious that various factors can contribute to make any static/non-static system more complex. For the considered structure, we indicate $Y_{TF}$ as a complexity factor in which certain components such as inhomogeneous energy density, anisotropic pressure and electric charge are participated under the presence of modified correction terms. In relevant field equations, five unknown functions $(\mu,P_{r},P_{t},X,Y)$ are to be found. Therefore, two more conditions are required for our further work and from them, one could be the disappering complexity factor condition (by putting $Y_{TF}=0$ in Eq.\eqref{f58}), as
\begin{align}\nonumber
\Pi&=\frac{H}{2}\left[\frac{1}{r^3}\int_{0}^{r}\tilde{r}^3\left(\frac{\mu^{(eff)}+Q_{0}^{(eff)}}{H}\right)'d\tilde{r}
-\frac{1}{H}(\Pi^{(D)}+Q_{1}^{(eff)}-Q_{2}^{(eff)})\right.\\\label{f62}
&-\left.\frac{1}{H}\left(f_{T}+\frac{1}{2}Rf_{Q}\right)\left(\Pi-\frac{s^2}{4\pi r^4}\right)+\frac{s^2}{4\pi Hr^4}-\frac{1}{4\pi}\varphi_{3}^{(D)}\right].
\end{align}
In modified gravity theory, Eq.\eqref{f62} could be regarded as a non-local equation of state. In next two subsections, different models of stellar objects are discussed.

\subsection{The Gokhroo and Mehra Ans\"{a}tz}

Gokhroo and Mehra \cite{gokhroo1994anisotropic} supposed a particular form of the energy density of a stellar object and analyzed the formation of its internal structure. So, we proceed their assumption in this case as
\begin{equation}\label{f63}
\frac{\mu^{(eff)}}{H}=\mu_{o}\left(1-\frac{Kr^2}{r_{\Sigma}^2}\right),
\end{equation}
where $K\in(0,1)$. One may write the following by using Eq.\eqref{f63} in Eq.\eqref{f21} as
\begin{equation}\label{f64}
m(r)=\frac{r}{8}+\frac{s^2}{2r}+\beta r^3\left(-\frac{1}{3}+\frac{Kr^2}{5r_{\Sigma}^2}\right)-4\pi \int_{0}^{r}\frac{\tilde{r}^2Q_{0}^{(eff)}}{H}d\tilde{r},
\end{equation}
and $\beta=4\pi\mu_{0}$. Also, by combining Eqs.\eqref{f20} and \eqref{f64}, we get
\begin{equation}\label{f65}
\frac{1}{Y}=\sqrt{\frac{2\beta r^3}{ r}\left(\frac{1}{3}-\frac{Kr^2}{5r_{\Sigma}^2}\right)
+\frac{8\pi}{r}\int_{0}^{r}\frac{\tilde{r}^2Q_{0}^{(eff)}}{H}d\tilde{r}}.
\end{equation}
Futher, from field equations \eqref{f14} and \eqref{f15} we have
\begin{eqnarray}\label{f66}
\frac{8\pi}{H}\left[\Pi^{(eff)}+Q_{1}^{(eff)}-Q_{2}^{(eff)}\right]=\frac{1}{Y^2}\left[-\frac{X'}{rX}
-\frac{X'Y'}{XY}-\frac{1}{r^2}+\frac{X''}{X}-\frac{Y'}{rY}\right].
\end{eqnarray}
We propose some new variables which can be useful as
\begin{equation}\label{f67}
X^2=e^{\int(2z(r)-2/r)dr},\quad Y^{-2}=y(r),
\end{equation}
and putting back these variables in Eq.\eqref{f66}, we get
\begin{eqnarray}\label{f68}
y'+y\left[\frac{2z'}{z}+2z+\frac{4}{r^2z}-\frac{6}{r}\right]&=&\frac{16\pi}{Hz}\left(\Pi^{(eff)}+Q_{1}^{(eff)}-Q_{2}^{(eff)}\right),
\end{eqnarray}
which looks to be in the proposed form by Ricatti. In $f(R,T,Q)$ theory, the line element \eqref{f12} appears in terms of new variables after the above expression has been integrated as
\begin{align}\nonumber
ds^2&=e^{\int(2z(r)-2/r)dr}dt^2\\\nonumber
&-\frac{z^2(r)e^{\int \left(\frac{4}{r^2z(r)}+2z(r)\right)dr}}{r^6\left[16\pi\int\left\{\frac{z\left(\Pi^{(eff)}+Q_{1}^{(eff)}-Q_{2}^{(eff)}\right)e^{\int \left(\frac{4}{r^2z(r)}+2z(r)\right)dr}}{Hr^6}\right\} dr+C\right]}dr^2\\\label{f69}
&-r^2(d\theta^2+\alpha^2 dz^2),
\end{align}
where $C$ presents as a constant of integration. In addition, the field equations \eqref{f13}-\eqref{f15} can also be expressed in terms of variables \eqref{f67} as
\begin{align}\label{f70}
\frac{4\pi}{H}\left[\mu^{(eff)}+Q_{0}^{(eff)}\right]&=\frac{1}{2r^2}-\frac{m'}{r^2}+\frac{s}{r^3}
\left(s'-\frac{s}{2r}\right),
\\\label{f71}
\frac{4\pi}{H}\left[P^{(eff)}_{r}+Q_{1}^{(eff)}\right]&=\left(\frac{1}{2}-\frac{4m}{r}\right)
\left(\frac{1}{4r^2}-\frac{z}{2r}\right)+\frac{s^2}{r^3}\left(\frac{1}{2r}-z\right),
\\\nonumber
\frac{4\pi}{H}\left[P^{(eff)}_{t}+Q_{0}^{(eff)}\right]&=\left(-\frac{1}{8}+\frac{m}{r}\right)\left(z'+z^2-\frac{3z}{2r}
+\frac{1}{r^2}\right)+\frac{m'z}{2r}\\\label{f72}
&-\frac{s^2}{2r^2}\left(z'+z^2-\frac{2z}{r}+\frac{1}{r^2}+\frac{s'z}{s}\right).
\end{align}
In cylindrical structure, Eqs.\eqref{f70}-\eqref{f72} may be helpful in understanding its mysterious and highly fascinating properties. Di Prisco \emph{et al.} \cite{di2011expansion} done the same work in the background of GR and calculated their corresponding solutions, whereas Sharif and Yousaf \cite{sharif2012shearfree,sharif2012expansion} extended this for spherical and cylindrical structures in modified framework. Also, Yousaf \cite{yousaf2017spherical} and Bhatti \cite{bhatti2016shear} found the more generalized results for Einstein-$\Lambda$ gravity. The conditions \eqref{f23} calculated at the boundary $\Sigma$ must satisfied by the obtained solutions to prevent the effective material variables from being of a singular nature.

\subsection{The Polytropic Equation With Zero Complexity Factor}

The polytropical equation of state is a key to analyze any self-gravitating system. To explain the effects of modified corrections in our model, we have to solve a system of equations for which it is required to combine this equation of state with the disappearing complexity factor condition. We will discuss here two individual cases of polytropes in charged cylindrical star, of which the first is
\begin{equation}\label{f73}
P^{(eff)}_{r}=\omega[\mu^{(eff)}]^{\rho}=\omega[\mu^{(eff)}]^{(1+1/n)}; \quad Y_{TF}=0,
\end{equation}
where $\omega$ and $\rho$ denote the polytropic constant and polytropic exponent. Also, $n$ is termed as polytropic index.

In order to solve the system of equations easily, we are introducing few variables to write these equations in dimensionless form as
\begin{equation}\label{f74}
\sigma=P^{(eff)}_{rc}/\mu^{(eff)}_{c},\quad r=\eta/A,\quad A^2=4\pi\mu^{(eff)}_{c}/\sigma(n+1),
\end{equation}
\begin{equation}\label{f75}
\phi^n=\mu^{(eff)}/\mu^{(eff)}_{c},\quad \nu(\eta)=m(r)A^3/(4\pi\mu^{(eff)}_{c}),
\end{equation}
where $\mu_{c}^{(eff)}$ and $P_{rc}^{(eff)}$ show that these quantities are calculated at the center. Also, at the boundary $r=r_{\Sigma}$, we have $\eta=\eta_{\Sigma}$ and thus $\phi(\eta_{\Sigma})=0$. After utilizing Eqs. \eqref{f74} and \eqref{f75} in TOV equation \eqref{f19}, it becomes
\begin{align}\nonumber
&\frac{\eta^2}{1+\sigma\phi}\left[1-\frac{8\nu\sigma(n+1)}{\eta}+\frac{4s^2A^2}{\eta^2}\right]\frac{d\phi}{d\eta}
+\frac{2\phi^{-n}\eta}{P^{(eff)}_{rc}(n+1)(1+\sigma\phi)}\left(\Pi^{(eff)}\right.\\\nonumber
&+\left.Q_{1}^{(eff)}-Q_{2}^{(eff)}\right)\left[1-\frac{8\nu\sigma(n+1)}{\eta}+\frac{4s^2A^2}{\eta^2}\right]
+4\left(1+\frac{Q_{5}^{(eff)}\phi^{-n}}{\mu_{c}^{(eff)}(1+\sigma\phi)}\right)\\\nonumber
&\times\left[\frac{\sigma\eta^3\phi^{n+1}}{H}+\frac{\eta^3Q_{1}^{(eff)}}{H\mu_{c}^{(eff)}}-\nu+\frac{A^2\eta}{32\pi\mu_{c}^{(eff)}}
+\frac{s^2A^4}{8\pi\eta A\mu_{c}^{(eff)}}\right]=\frac{\eta^2\phi^{-n}}{AP_{rc}^{(eff)}(n+1)}\\\label{f76}
&\times\frac{1}{(1+\sigma\phi)}\left[1-\frac{8\nu\sigma(n+1)}{\eta}+\frac{4s^2A^2}{\eta^2}\right]
\left\{Y^2ZH+\frac{A}{H}\frac{dH}{d\eta}\left(\phi^{n+1}P_{rc}^{(eff)}+Q_{1}^{(eff)}\right)\right\}.
\end{align}
By taking derivative of $\nu(\eta)$ and combining it with Eq.\eqref{f70}, we have
\begin{equation}\label{f77}
\frac{d\nu}{d\eta}=\frac{A^2}{32\pi\mu_{c}^{(eff)}}-\frac{\eta^2}{H}\left(\phi^n+\frac{Q_{0}^{(eff)}}
{\mu_{c}^{(eff)}}\right)+\frac{sA^4}{8\pi\eta\mu_{c}^{(eff)}}\left(2\frac{ds}{d\eta}-\frac{s}{\eta}\right).
\end{equation}
It should be mentioned that two ordinary differential equations (ODE's) \eqref{f76} and \eqref{f77} consist of three unknown functions $\phi,~\nu$ and $\Pi$. We require one more equation to have a unique solution of this system and thus the dimensionless form of vanishing complexity factor condition \eqref{f62} becomes
\begin{align}\nonumber
&\frac{6\Pi}{n\mu^{(eff)}_{c}}+\frac{2\eta}{n\mu^{(eff)}_{c}}\frac{d\Pi}{d\eta}=\phi^{n-1}\eta\frac{d\phi}{d\eta}
+\frac{2\Pi\eta}{Hn\mu^{(eff)}_{c}}\frac{dH}{d\eta}-\frac{3}{n\mu^{(eff)}_{c}}\left[\left(\Pi-\frac{s^2A^4}{4\pi\eta^4}\right)\right.\\\nonumber
&\times\left(f_{T}+\frac{1}{2}Rf_{Q}\right)-\left.\frac{s^2A^4}{4\pi\eta^4}+\Pi^{(D)}+Q_{1}^{(eff)}-Q_{2}^{(eff)}+\frac{H}{4\pi}\varphi_{3}^{(D)}\right]
+\frac{H\eta\phi^n}{n}\frac{d}{d\eta}\left(\frac{1}{H}\right)\\\nonumber
&+\frac{H\eta}{n\mu^{(eff)}_{c}}\left[\frac{d}{d\eta}\left(\frac{Q^{(eff)}_{0}}{H}\right)-\frac{d}{d\eta}\left\{\frac{1}{H}\left(\Pi^{(D)}+Q_{1}^{(eff)}
-Q_{2}^{(eff)}\right)\right\}-\frac{d}{d\eta}\left\{\frac{1}{H}\right.\right.\\\label{f78}
&\times\left.\left.\left(\Pi-\frac{s^2A^4}{4\pi\eta^4}\right)\left(f_{T}+\frac{1}{2}Rf_{Q}\right)\right\}+\frac{sA^4}{2\pi H\eta^4}\frac{ds}{d\eta}-\frac{s^2A^4}{4\pi H^2\eta^4}\frac{dH}{d\eta}-\frac{s^2A^4}{\pi H \eta^5}-\frac{1}{4\pi}\frac{d\varphi_{3}^{(D)}}{d\eta}\right].
\end{align}
We finally get a system of three ODE's \eqref{f76},~\eqref{f77} and \eqref{f78}. This system provides us a unique solution by putting arbitrary values of parameters $\sigma$ and $n$. One can study some properties of stellar object such as mass and density with the help of solution of the above system which may obtain for particular values of $\sigma$ and $n$.

We are now ready to consider the second case of the polytropic equation of state as $P^{(eff)}_{r}=\omega[\mu_{b}^{(eff)}]^{\rho}=\omega[\mu_{b}^{(eff)}]^{(1+1/n)}$, where $\mu_{b}$ indicate the baryonic (rest) mass density. Therefore, Eqs.\eqref{f76} and \eqref{f78} can be rewritten in this case as
\begin{align}\nonumber
&\frac{\eta^2}{1+\sigma\phi_{b}}\left[1-\frac{8\nu\sigma(n+1)}{\eta}+\frac{4s^2A^2}{\eta^2}\right]\frac{d\phi_{b}}{d\eta}
+\frac{2\phi_{b}^{-n}\eta}{P^{(eff)}_{rc}(n+1)(1+\sigma\phi_{b})}\left(\Pi^{(eff)}\right.\\\nonumber
&+\left.Q_{1}^{(eff)}-Q_{2}^{(eff)}\right)\left[1-\frac{8\nu\sigma(n+1)}{\eta}+\frac{4s^2A^2}{\eta^2}\right]
+4\left(1+\frac{Q_{5}^{(eff)}\phi_{b}^{-n}}{\mu_{bc}^{(eff)}(1+\sigma\phi_{b})}\right)\\\nonumber
&\times\left[\frac{\sigma\eta^3\phi_{b}^{n+1}}{H}+\frac{\eta^3Q_{1}^{(eff)}}{H\mu_{bc}^{(eff)}}-\nu+\frac{A^2\eta}{32\pi\mu_{bc}^{(eff)}}
+\frac{s^2A^4}{8\pi\eta A\mu_{bc}^{(eff)}}\right]=\frac{\eta^2\phi_{b}^{-n}}{AP_{rc}^{(eff)}(n+1)}\\\label{f79}
&\times\frac{1}{(1+\sigma\phi_{b})}\left[1-\frac{8\nu\sigma(n+1)}{\eta}+\frac{4s^2A^2}{\eta^2}\right]
\left\{Y^2ZH+\frac{A}{H}\frac{dH}{d\eta}\left(\phi_{b}^{n+1}P_{rc}^{(eff)}+Q_{1}^{(eff)}\right)\right\},
\end{align}
and
\begin{align}\nonumber
&\frac{6\Pi}{n\mu^{(eff)}_{bc}}+\frac{2\eta}{n\mu^{(eff)}_{bc}}\frac{d\Pi}{d\eta}=\phi_{b}^{n-1}\eta\left[1+K(n+1)
\left(\mu_{bc}^{(eff)}\right)^{1/n}\phi_{b}\right]\frac{d\phi_{b}}{d\eta}+\frac{2\Pi\eta}{Hn\mu^{(eff)}_{bc}}\\\nonumber
&\times\frac{dH}{d\eta}-\frac{3}{n\mu^{(eff)}_{bc}}\left[\left(\Pi-\frac{s^2A^4}{4\pi\eta^4}\right)\left(f_{T}
+\frac{1}{2}Rf_{Q}\right)-\frac{s^2A^4}{4\pi\eta^4}+\Pi^{(D)}+Q_{1}^{(eff)}\right.\\\nonumber
&-\left.Q_{2}^{(eff)}+\frac{H}{4\pi}\varphi_{3}^{(D)}\right]+\frac{H\eta\phi_{b}^n}{n}\frac{d}{d\eta}\left(\frac{1}{H}\right)+\frac{H\eta}{n\mu^{(eff)}_{bc}}
\left[\frac{d}{d\eta}\left(\frac{Q^{(eff)}_{0}}{H}\right)-\frac{d}{d\eta}\left\{\frac{1}{H}\left(\Pi^{(D)}\right.\right.\right.\\\nonumber
&+\left.\left.Q_{1}^{(eff)}-Q_{2}^{(eff)}\right)\right\}-\frac{d}{d\eta}\left\{\frac{1}{H}\left(\Pi-\frac{s^2A^4}{4\pi\eta^4}\right)
\left(f_{T}+\frac{1}{2}Rf_{Q}\right)\right\}+\frac{sA^4}{2\pi H\eta^4}\frac{ds}{d\eta}\\\label{f80}
&-\left.\frac{s^2A^4}{4\pi H^2\eta^4}\frac{dH}{d\eta}-\frac{s^2A^4}{\pi H \eta^5}-\frac{1}{4\pi}\frac{d\varphi_{3}^{(D)}}{d\eta}\right],
\end{align}
where $\phi_{b}^n=\mu_{b}^{(eff)}/\mu_{bc}^{(eff)}$.
\section{Conclusions}

The basic objective in doing this work is to analyze the self-gravitating system having cylindrical symmetry under $f(R,T,Q)$ corrections and for this reason, we considered anisotropic static charged geometry. The field equations containing $f(R,T,Q)$ corrections, C-energy and Tolman mass function are derived, and their relations with Conformal tensor are also developed. The tensors $X_{\rho\gamma},~Y_{\rho\gamma}$ and $Z_{\rho\gamma}$ are found from proper orthogonal splitting of the Riemann tensor and then received five modified scalar functions from them. After this, the effects of these scalars on the sustainability and evolution of cylindrical relativistic fluid were studied. Herrera \cite{herrera2018new} suggested the idea to find complexity of spherical star by preceding the assumption that the system with homogeneous energy density and isotropic pressure is simplest, i.e., there is no complexity. The corresponding modified scalar $Y_{TF}$ is therefore known as the complexity factor in this case. Thus, we are focused on some significant changes in $f(R,T,Q)$ theory.\\

(i) The effective scalar $Y_{TF}$ includes the effects of electric charge as we are studying charged cylindrical object.\\

(ii) For this charged case, the system becomes more complex (we can see from Eq.\eqref{f58}), because $Y_{TF}$ also contains the effective charge along with effective inhomogeneous energy density and effective pressure anisotropy.\\

(iii) The Tolman mass in $f(R,T,Q)$ theory could be measured via this scalar $Y_{TF}$ in terms of above effective physical quantities.\\

After calculating the Darmois junction conditions on the boundary $\Sigma$ in modified gravity, we briefly discussed the electric and magnetic parts of Weyl tensor. In addition, we have examined two different applications of stellar systems by adding modified corrections through vanishing of complexity factor \eqref{f58}. In first of them, we modified the form of energy density which was suggested by Gokhroo and Mehra to analyze the characteristics of stellar structures. In second example, we discussed the equation of state for polytropic substance and also added certain variables to make some equations dimensionless. For some realistic constraints, the obtained system of modified equations offer a solution that allows us to better understand the system. To get all our results in GR with charge free case, one can substitute $f(R,T,Q)=R$ and $L_{EM}=0$ in action function \eqref{f1}.

\section*{Appendix A}

The effective quantities appearing in field equations \eqref{f13}-\eqref{f15} are
\begin{align}\nonumber
\mu^{(eff)}&=\mu\left[1+2f_{T}+f_{Q}\left(\frac{3X'Y'}{2XY^3}-\frac{3X'}{rXY^2}+\frac{1}{2}R-\frac{3X''}{2XY^2}\right)-f'_{Q}\left(\frac{Y'}{2Y^3}\right.\right.\\\nonumber
&-\left.\left.\frac{1}{rY^2}\right)+\frac{f''_{Q}}{2Y^2}\right]
+\mu'\left[-f_{Q}\left(\frac{Y'}{2Y^3}-\frac{1}{rY^2}\right)+\frac{f'_{Q}}{Y^2}\right]
+\frac{\mu''f_{Q}}{2Y^2}+P_{r}\left[f_{Q}\right.\\\nonumber
&\times\left.\left(\frac{Y'}{rY^3}-\frac{1}{r^2Y^2}+\frac{X''}{2XY^2}-\frac{X'Y'}{2XY^3}\right)-f'_{Q}\left(\frac{2}{rY^2}
-\frac{Y'}{2Y^3}\right)-\frac{f''_{Q}}{2Y^2}\right]\\\nonumber
&-P'_{r}\left[f_{Q}\left(\frac{2}{rY^2}-\frac{Y'}{2Y^3}\right)
+\frac{f'_{Q}}{Y^2}\right]-\frac{P''_{r}f_{Q}}{2Y^2}+P_{t}\left[f_{Q}\left(\frac{1}{r^2Y^2}+\frac{X'}{rXY^2}\right.\right.\\\label{f81}
&-\left.\left.\frac{Y'}{rY^3}\right)+\frac{f'_{Q}}{rY^2}\right]+\frac{P'_{t}f_{Q}}{rY^2}+\frac{R}{2}\left(\frac{f}{R}-f_{R}\right)-f'_{R}\left(\frac{Y'}{Y^3}-\frac{2}{rY^2}\right)+\frac{f''_{R}}{Y^2},
\\\nonumber
P_{r}^{(eff)}&=\mu\left[-f_{T}+f_{Q}\left(\frac{X''}{2XY^2}+\frac{X'}{rXY^2}-\frac{X'Y'}{2XY^3}\right)
-\frac{f'_{Q}X'}{2XY^2}\right]-\frac{\mu'X'f_{Q}}{2XY^2}\\\nonumber
&+P_{r}\left[1+f_{T}+f_{Q}\left(\frac{2X'}{rXY^2}+\frac{1}{2}R+\frac{1}{r^2Y^2}+\frac{3X'Y'}{2XY^3}+\frac{3Y'}{rY^3}-\frac{3X''}{2XY^2}\right)\right.\\\nonumber
&+\left.f'_{Q}\left(\frac{1}{rY^2}+\frac{X'}{2XY^2}\right)\right]+P'_{r}\left[f_{Q}\left(\frac{1}{rY^2}+\frac{X'}{2XY^2}\right)\right]
+P_{t}\left[f_{Q}\left(\frac{Y'}{rY^3}\right.\right.\\\nonumber
&-\left.\left.\frac{1}{r^2Y^2}-\frac{X'}{rXY^2}\right)+\frac{3f'_{Q}}{2rY^2}\right]+\frac{P'_{t}f_{Q}}{rY^2}
+\frac{R}{2}\left(f_{R}-\frac{f}{R}\right)-f'_{R}\\\label{f82}
&\times\left(\frac{2}{rY^2}+\frac{X'}{XY^2}\right),
\\\nonumber
P_{t}^{(eff)}&=\mu\left[-f_{T}+f_{Q}\left(\frac{X''}{2XY^2}+\frac{X'}{rXY^2}-\frac{X'Y'}{2XY^3}\right)
+\frac{f'_{Q}X'}{2XY^2}\right]+\frac{\mu'X'f_{Q}}{2XY^2}\\\nonumber
&+P_{r}\left[f_{Q}\left(\frac{X''}{2XY^2}+\frac{X'}{rXY^2}-\frac{X'Y'}{2XY^3}\right)
+f'_{Q}\left(\frac{1}{rY^2}+\frac{X'}{XY^2}-\frac{Y'}{2Y^3}\right)\right.\\\nonumber
&+\left.\frac{f''_{Q}}{2Y^2}\right]+P'_{r}\left[f_{Q}\left(\frac{1}{rY^2}+\frac{X'}{XY^2}-\frac{Y'}{2Y^3}\right)+\frac{f'_{Q}}{Y^2}\right]
+\frac{P''_{r}f_{Q}}{2Y^2}+P_{t}\left[1+f_{T}\right.\\\nonumber
&+\left.\left.f_{Q}\left(\frac{2Y'}{rY^3}+\frac{1}{2}R-\frac{2}{r^2Y^2}-\frac{2X'}{rXY^2}\right)
-f'_{Q}\left(\frac{Y'}{2Y^3}-\frac{X'}{2XY^2}\right)+\frac{f''_{Q}}{2Y^2}\right]\right.\\\nonumber
&+P'_{t}\left[-f_{Q}\left(\frac{Y'}{2Y^3}-\frac{X'}{2XY^2}\right)
+\frac{f'_{Q}}{Y^2}\right]+\frac{P''_{t}f_{Q}}{2Y^2}+\frac{R}{2}\left(f_{R}-\frac{f}{R}\right)-f'_{R}\\\label{f83}
&\times\left(\frac{1}{rY^2}-\frac{Y'}{Y^3}+\frac{X'}{XY^2}\right)-\frac{f''_{R}}{Y^2},
\\\nonumber
Q_{0}^{(eff)}&=\left(1+\frac{1}{2}Rf_{Q}+f_{T}\right)\frac{s^2}{8\pi r^4}-\frac{s^2f_{T}}{2r^4}
+f''_{Q}\left(\frac{s^2}{8\pi r^4X^2}+\frac{s^2}{8\pi r^4Y^2}\right)\\\nonumber
&+f'_{Q}\left(\frac{s^2X'}{8\pi r^4XY^2}-\frac{s^2}{2\pi r^5Y^2}+\frac{ss'}{2\pi r^4Y^2}-\frac{s^2Y'}{8\pi r^4Y^3}-\frac{s^2Y'}{8\pi r^4X^2Y}\right.\\\nonumber
&+\left.\frac{ss'}{2\pi r^4X^2}-\frac{s^2}{2\pi r^5X^2}\right)+f_{Q}\left(\frac{ss'X'}{4\pi r^4XY^2}-\frac{ss'}{\pi r^5Y^2}+\frac{3s^2X'}{8\pi r^5XY^2}\right.\\\nonumber
&+\frac{s'^2}{4\pi r^4Y^2}+\frac{ss''}{4\pi r^4Y^2}-\frac{ss'Y'}{4\pi r^4Y^3}+\frac{3s^2}{4\pi r^6Y^2}+\frac{s^2Y'}{4\pi r^5Y^3}+\frac{s'^2}{4\pi r^4X^2}\\\label{f84}
&+\left.\frac{ss''}{4\pi r^4X^2}-\frac{ss'}{\pi r^5X^2}-\frac{ss'Y'}{4\pi r^4X^2Y}+\frac{s^2X''}{4\pi r^4X^3}
-\frac{s^2X'Y'}{4\pi r^4X^3Y}\right),
\\\nonumber
Q_{1}^{(eff)}&=-\left(1+\frac{1}{2}Rf_{Q}+f_{T}\right)\frac{s^2}{8\pi r^4}+\frac{s^2f_{T}}{2r^4}
-\frac{f'_{Q}s^2X'}{8\pi r^4XY^2}+f_{Q}\left(\frac{s^2X''}{4\pi r^4XY^2}\right.\\\nonumber
&-\frac{s^2X'Y'}{4\pi r^4XY^3}+\frac{ss'Y'}{4\pi r^4Y^3}-\frac{ss'X'}{4\pi r^4XY^2}+\frac{s^2X'}{8\pi r^5XY^2}-\frac{s^2}{4\pi r^6Y^2}\\\label{f85}
&-\left.\frac{3s^2Y'}{4\pi r^5Y^3}\right),
\\\nonumber
Q_{2}^{(eff)}&=\left(1+\frac{1}{2}Rf_{Q}+f_{T}\right)\frac{s^2}{8\pi r^4}+\frac{s^2f_{T}}{2r^4}-\frac{f'_{Q}s^2}{8\pi r^5Y^2}+f_{Q}\left(\frac{s^2Y'}{4\pi r^5Y^3}\right.\\\label{f86}
&-\left.\frac{ss'}{4\pi r^5Y^2}-\frac{3s^2X'}{8\pi r^5XY^3}+\frac{s^2}{4\pi r^6Y^2}\right).
\end{align}

The value of $Z$ which is arisen in Eq.\eqref{f17} is
\begin{align}\nonumber
Z &= \frac{2}{\left(2+Rf_{Q}+2f_{T}\right)}\left[\frac{f'_{Q}}{Y^2}\left(P_{r}-\frac{s^2}{8\pi r^4}\right)\left(\frac{Y'}{rY}+\frac{X'Y'}{2XY}+\frac{1}{2r^2}+\frac{X'}{rX}
\right)+\frac{1}{2Y^2}\right.\\\nonumber
&\times\left(P_{r}-\frac{s^2}{8\pi r^4}\right)\left\{f_{Q}\left(\frac{X''Y'}{XY}+\frac{2Y''}{rY}-\frac{6Y'^2}{rY^2}-\frac{9X'Y'^2}{XY^2}-\frac{X'^2Y'}{X^2Y}+\frac{X'Y''}{XY}\right.\right.\\\nonumber
&-\left.\left.\frac{4Y'}{r^2Y}
-\frac{2X'}{r^2X}+\frac{2X''}{rX}-\frac{4X'Y'}{rXY}-\frac{2X'^2}{rX^2}-\frac{2}{r^3}\right)+2f'_{T}Y^2\right\}+\frac{1}{2Y^2}\left(P'_{r}\right.\\\nonumber
&-\left.\frac{ss'}{4\pi r^4}+\frac{s^2}{2\pi r^5}\right)\left\{f_{Q}\left(\frac{X'Y'}{XY}
+\frac{2Y'}{rY}-\frac{X''}{X}\right)+f_{T}Y^2\right\}-\frac{3f_{T}}{2}\left(\mu'-\frac{ss'}{r^4}\right.\\\nonumber
&\left.+\frac{2s^2}{r^5}\right)- f'_{T}\left(\mu -\frac{s^2}{2r^4}\right)-\frac{f_{Q}}{XY^2}\left(\mu'+\frac{ss'}{4\pi r^4}-\frac{s^2}{2\pi r^5}\right)\left(\frac{Y'}{r}-\frac{X'Y'}{2Y}+\frac{X''}{2}\right)\\\nonumber
&-\frac{1}{r^2Y^2}\left(P'_{t}+\frac{ss'}{4\pi r^2}-\frac{s^2}{4\pi r^3}\right)\left\{f_{Q}\left(\frac{rY'}{Y}-1-\frac{rX'}{X}\right)-f_{T}r^2Y^2\right\}+\left(\frac{1}{r^2}\right.\\\label{f87}
&+\left.\left.\frac{2X'}{rX}\right)\left\{f_{Q}\left(\mu'-\frac{ss'}{r^4}+\frac{2s^2}{r^5}\right)+ f'_{Q}\left(\mu-\frac{s^2}{2r^4}\right)\right\}\right].
\end{align}

The modified correction terms $F_{0}$ and $F_{1}$ in Eq.\eqref{f23} are given as
\begin{eqnarray}\nonumber
F_{0}&=&\tilde{f}_{T}\left(P_{r}-\mu\right)+\tilde{f}_{Q}\left[\mu\left(\frac{X''}{2XY^2}+\frac{X'}{rXY^2}
-\frac{X'Y'}{2XY^3}\right)-\frac{\mu'X'}{2XY^2}\right.\\\nonumber
&+&P_{r}\left(\frac{2X'}{rXY^2}+\frac{1}{2}R-\frac{3X''}{2XY^2}
+\frac{1}{r^2Y^2}+\frac{3X'Y'}{2XY^3}+\frac{3Y'}{rY^3}\right)+P'_{r}\left(\frac{X'}{2XY^2}\right.\\\label{f88}
&+&\left.\left.\frac{1}{rY^2}\right)+P_{t}\left(\frac{Y'}{rY^3}-\frac{1}{r^2Y^2}-\frac{X'}{rXY^2}\right)
+\frac{P'_{t}}{rY^2}\right]+\frac{R}{2}\left(\tilde{f}_{R}-\frac{f}{R}\right),
\\\nonumber
F_{1}&=&-\left(\frac{1}{2}R\tilde{f}_{Q}+\tilde{f}_{T}\right)\frac{s^2}{8\pi r^4}+\frac{s^2\tilde{f}_{T}}{2r^4}
+\tilde{f}_{Q}\left(\frac{s^2X''}{4\pi r^4XY^2}-\frac{s^2X'Y'}{4\pi r^4XY^3}\right.\\\nonumber
&+&\frac{ss'Y'}{4\pi r^4Y^3}-\frac{ss'X'}{4\pi r^4XY^2}+\frac{s^2X'}{8\pi r^5XY^2}-\frac{s^2}{4\pi r^6Y^2}+\frac{s^2X'Y'}{r^4XY^3}-\frac{s^2X''}{r^4XY^2}\\\label{f89}
&-&\left.\frac{3s^2Y'}{4\pi r^5Y^3}\right).
\end{eqnarray}

\section*{Appendix B}

The effective terms $\varphi_{1}^{(D)},~\varphi_{2}^{(D)}$ and $\varphi_{\delta\gamma}^{(D)}$ appeared in structure scalars \eqref{f53}, \eqref{f55} and \eqref{f56} are given as
\begin{eqnarray}\nonumber
\varphi_{1}^{(D)}&=&\frac{4\pi}{H}\left[\left\{h^{\gamma}_{\nu}\Box(f_{Q}X^{\nu}_{\gamma})-2h^{\gamma}_{\nu}\nabla^{\nu}\nabla_{\gamma}f_{R}
-h^{\gamma}_{\nu}\nabla_{\pi}\nabla_{\gamma}(f_{Q}X^{\pi\nu})\right.\right.\\\nonumber
&-&\left.h^{\gamma}_{\nu}\nabla_{\pi}\nabla^{\nu}(f_{Q}X^{\pi}_{\gamma})\right\}
+2f_{Q}\left(R^{\nu}_{\pi}h^{\pi}_{\nu}+R_{\pi\gamma}h^{\gamma\pi}\right)\left\{\frac{1}{3}\left(\Pi-\frac{s^2}{4\pi r^4}\right)\right.\\\nonumber
&-&\left.\left.\left(P+\frac{s^2}{24\pi r^4}\right)\right\}\right]
+\frac{8\pi}{H}\left[\left\{\frac{R}{2}\left(\frac{f}{R}-f_{R}\right)+\left(\mu-\frac{s^2}{2r^4}\right) f_{T}\right.\right.\\\nonumber
&-&\left.\frac{1}{2}\nabla_{\pi}\nabla_{\rho}(f_{Q}X^{\pi\rho})\right\}+2Rf_{Q}\left\{\left(P+\frac{s^2}{24\pi r^4}\right)-\frac{1}{3}\left(\Pi-\frac{s^2}{4\pi r^4}\right)\right\}\\\nonumber
&+&\nabla_{\pi}\nabla_{\nu}(f_{Q}X^{\pi\nu})-\frac{1}{2}\Box\{f_{Q}(\mu-3P)\}+2g^{\nu\gamma}(f_{Q}R^{\pi\rho}+f_{T}g^{\pi\rho})\\\label{f90}
&\times&\left.\frac{\partial^2L_{m}}{\partial g^{\nu\gamma}\partial g^{\pi\rho}}\right],
\\\nonumber
\varphi_{2}^{(D)}&=&-\frac{8\pi}{H}\left\{\frac{R}{2}\left(\frac{f}{R}-f_{R}\right)+\left(\mu-\frac{s^2}{2r^4}\right) f_{T}-\frac{1}{2}\nabla_{\pi}\nabla_{\rho}(f_{Q}X^{\pi\rho})\right\}+\frac{4\pi}{H}\\\nonumber
&\times&\left[-\frac{1}{2}\left\{\Box(f_{Q}X)+3v^{\alpha}v_{\delta}\Box(f_{Q}X^{\delta}_{\alpha})-v^{\delta}v^{\alpha}\Box(f_{Q}X_{\delta\alpha})
\right\}+\left(\Box f_{R}+3v^{\alpha}v_{\delta}\right.\right.\\\nonumber
&\times&\left.\nabla^{\delta}\nabla_{\alpha}f_{R}-v^{\delta}v^{\alpha}\nabla_{\delta}\nabla_{\alpha}f_{R}\right)
-2f_{Q}R^{\alpha}_{\pi}\left\{\left(\Pi-\frac{s^2}{4\pi r^4}\right)\left(w^{\pi}w_{\alpha}
+\frac{1}{3}h^{\pi}_{\alpha}\right)\right.\\\nonumber
&-&\left.\left(P+\frac{s^2}{24\pi r^4}\right) h^{\pi}_{\alpha}\right\}-3f_{Q}\left(R_{\pi\alpha} v^{\pi}v^{\alpha}+R^{\delta}_{\pi}v^{\pi}v_{\delta}\right)\left(\mu+\frac{s^2}{8\pi r^4}\right)\\\nonumber
&+&\frac{1}{2}\{\nabla_{\pi}\nabla_{\delta}(f_{Q}X^{\pi\delta})+4v_{\delta}v^{\alpha}\nabla_{\pi}\nabla^{\delta}(f_{Q}X^{\pi}_{\alpha})
+\nabla_{\pi}\nabla_{\alpha}(f_{Q}X^{\pi\alpha})+4v_{\delta}v^{\alpha}\\\nonumber
&\times&\nabla_{\pi}\nabla_{\alpha}(f_{Q}X^{\pi\delta})-v_{\delta}v^{\alpha}\nabla_{\pi}\nabla_{\alpha}(f_{Q}X^{\delta\pi})
-v^{\delta}v^{\alpha}\nabla_{\pi}\nabla_{\delta}(f_{Q}X^{\pi}_{\alpha})-v_{\delta}v^{\alpha}\\\nonumber
&\times&\nabla_{\pi}\nabla^{\delta}(f_{Q}X^{\pi}_{\alpha})
-v^{\delta}v^{\alpha}\nabla_{\pi}\nabla_{\alpha}(f_{Q}X^{\pi}_{\delta})\}+2h^{\epsilon\alpha}(f_{Q}R^{\pi\nu}+f_{T}g^{\pi\nu})\\\nonumber
&\times&\left.\frac{\partial^2L_{m}}{\partial g^{\epsilon\alpha}\partial g^{\pi\nu}}\right]+\frac{8\pi}{H}\left[\frac{1}{2}\Box\{f_{Q}(\mu-3P)\}
-\nabla_{\pi}\nabla_{\rho}(f_{Q}X^{\pi\rho})+2f_{Q}R_{\pi\rho}X^{\pi\rho}\right.\\\label{f91}
&-&\left.2g^{\rho\xi}(f_{Q}R^{\pi\nu}+f_{T}g^{\pi\nu})\frac{\partial^2L_{m}}{\partial g^{\rho\xi}\partial g^{\pi\nu}}\right],
\\\nonumber
\varphi_{\delta\gamma}^{(D)}&=&-\frac{2\pi}{H}\left[h^{\lambda}_{\delta}h^{\mu}_{\gamma}\Box(f_{Q}X_{\lambda\mu})
-v_{\delta}v_{\gamma}v_{\beta}v^{\alpha}\Box(f_{Q}X^{\beta}_{\alpha})-\Box(f_{Q}X_{\delta\gamma})\right]+\frac{4\pi}{H}\\\nonumber
&\times&\left[(h^{\lambda}_{\delta}h^{\mu}_{\gamma}\nabla_{\mu}\nabla_{\lambda}f_{R}
-v_{\delta}v_{\gamma}v_{\beta}v^{\alpha}\nabla^{\beta}\nabla_{\alpha}f_{R}-\nabla_{\delta}\nabla_{\gamma}f_{R})-f_{Q}(R_{\lambda\pi}h^{\lambda}_{\delta}\right.\\\nonumber
&-&R_{\delta\pi})\left\{\left(\Pi-\frac{s^2}{4\pi r^4}\right)\left(w^{\pi}w_{\gamma}+\frac{1}{3}h^{\pi}_{\gamma}\right)-\left(P+\frac{s^2}{24\pi r^4}\right)h^{\pi}_{\gamma}\right\}+f_{Q}(R_{\pi\gamma}\\\nonumber
&-&R_{\pi\mu}h^{\mu}_{\gamma})\left\{\left(\Pi-\frac{s^2}{4\pi r^4}\right)\left(w^{\pi}w_{\delta}
+\frac{1}{3}h^{\pi}_{\delta}\right)-\left(P+\frac{s^2}{24\pi r^4}\right)h^{\pi}_{\delta}\right\}\\\nonumber
&+&\frac{1}{2}\{h^{\lambda}_{\delta}h^{\mu}_{\gamma}\nabla_{\pi}\nabla_{\lambda}(f_{Q}X^{\pi}_{\mu})
-\nabla_{\pi}\nabla_{\gamma}(f_{Q}X^{\pi}_{\delta})+h^{\lambda}_{\delta}h^{\mu}_{\gamma}\nabla_{\pi}\nabla_{\mu}(f_{Q}X^{\pi}_{\lambda})\\\nonumber
&-&\nabla_{\pi}\nabla_{\delta}(f_{Q}X^{\pi}_{\gamma})
-v_{\delta}v_{\gamma}v_{\beta}v^{\alpha}\nabla_{\pi}\nabla^{\beta}(f_{Q}X^{\pi}_{\alpha})
-v_{\delta}v_{\gamma}v_{\beta}v^{\alpha}\nabla_{\pi}\nabla_{\alpha}(f_{Q}X^{\pi\beta})\}\\\label{f92}
&+&\left.2(f_{Q}R^{\pi\nu}+f_{T}g^{\pi\nu})h^{\epsilon}_{\delta}\left\{h^{\mu}_{\gamma}\frac{\partial^2L_{m}}{\partial g^{\epsilon\mu}\partial g^{\pi\nu}}-\frac{\partial^2L_{m}}{\partial g^{\epsilon\gamma}\partial g^{\pi\nu}}\right\}\right].
\end{eqnarray}

\vspace{0.5cm}

%%%%%%%%%%%%%%%%%%%%%%%%
%%%  Acknowledgments
%%%%%%%%%%%%%%%%%%%%%%%%
{\bf Acknowledgments}

\vspace{0.25cm}

This work was supported by National Research Project for Universities (NRPU), Higher Education Commission, Pakistan
under the research project No. 8754/Punjab/NRPU/R\&D/HEC/2017.

\end{document}